\documentclass[twocolumn,floatfix]{aastex631}

\usepackage[utf8]{inputenc}
\usepackage{natbib}
\usepackage{amsmath}

\usepackage{listings}
\usepackage{jlcode}
\usepackage{hyperref}

\newcommand{\octo}{Octofitter}
\newcommand{\julia}{Julia}

\newcommand{\colwidth}{\columnwidth}

\usepackage{newunicodechar}
\newunicodechar{σ}{\ensuremath{\mathrm{\sigma}}}
\shorttitle{Octofitter}
\shortauthors{Thompson et al.}

\graphicspath{{./}{figures/}}

\begin{document}

\title{Octofitter: fast, flexible, and accurate orbit modelling to detect exoplanets}

\author[0000-0001-5684-4593]{William Thompson}
\affiliation{
    Department of Physics and Astronomy, University of Victoria, \\
    Victoria, BC V8P 5C2, Canada}

\author[0009-0007-5371-3548]{Jensen Lawrence}
\affiliation{
    National Research Council of Canada Herzberg,\\
    Victoria, BC, V9E 2E7, Canada}

\author[0000-0001-9582-4261]{Dori Blakely}
\affiliation{
    Department of Physics and Astronomy, University of Victoria, \\
    3800 Finnerty Rd,
    Victoria, BC V8P 5C2, Canada}
\affiliation{
    National Research Council of Canada Herzberg,\\
    Victoria, BC, V9E 2E7, Canada}

\author[0000-0002-4164-4182]{Christian Marois}
\affiliation{
    National Research Council of Canada Herzberg,\\
    Victoria, BC, V9E 2E7, Canada}

\author[0000-0003-0774-6502]{Jason Wang}
\affiliation{
    Center for Interdisciplinary Exploration and Research in Astrophysics (CIERA) and Department of Physics and Astronomy, Northwestern University, Evanston, IL
    60208, USA
}

\author[0000-0002-7218-2873]{Mosè Giordano}
\affiliation{
    University College London,\\
    Gower St, London WC1E 6BT, United Kingdom
}

\author[0000-0003-2630-8073]{Timothy Brandt}
\affiliation{
    Department of Physics, 
    University of California,\\
    Santa Barbara, USA
}

\author[0000-0002-6773-459X]{Doug Johnstone}
\affiliation{
    National Research Council of Canada Herzberg,\\
    5071 West Saanich Rd,
    Victoria, BC, V9E 2E7, Canada}
\affiliation{
    Department of Physics and Astronomy, University of Victoria, \\
    Victoria, BC V8P 5C2, Canada}
    
\author[0000-0003-2233-4821]{Jean-Baptiste Ruffio}
\affiliation{
    Department of Astronomy, 
    California Institute of Technology,
    Pasadena, CA 91125, USA
}


\author{S. Mark Ammons} 
\affiliation{
    Lawrence Livermore National Laboratory
}

\author[0000-0003-4909-256X]{Katie A. Crotts} 
\affiliation{
    Department of Physics and Astronomy, University of Victoria, \\
    3800 Finnerty Rd,
    Victoria, BC V8P 5C2, Canada}

\author[0000-0001-5173-2947]{Clarissa R. Do \'O}
\affiliation{
    Center for Astrophysics and Space Sciences,
    University of California,
    San Diego, La Jolla, CA 92093, USA
}

\author[0000-0003-4636-6676]{Eileen C. Gonzales} 
\altaffiliation{51 Pegasi b Fellow}
\affiliation{Department of Astronomy and Carl Sagan Institute, Cornell University, 122 Sciences Drive, Ithaca, NY 14853, USA}

\author[0000-0002-7670-670X]{Malena Rice} 
\affiliation{Department of Astronomy, Yale University,\\New Haven, CT 06511, USA}


\begin{abstract}

%
As next-generation imaging instruments and interferometers search for planets closer to their stars, they must contend with increasing orbital motion and longer integration times.
These compounding effects make it difficult to detect faint planets but also present an opportunity. Increased orbital motion makes it possible to move the search for planets into the orbital domain, where direct images can be freely combined with radial velocity and proper motion anomaly, even without a confirmed detection in any single epoch. 
In this paper, we present a fast and differentiable multi-method orbit modelling and planet detection code called Octofitter.
This code is designed to be highly modular and allows users to easily adjust priors, change parameterizations, and specify arbitrary function relations between the parameters of one or more planets. Octofitter further supplies tools for examining model outputs including prior and posterior predictive checks and simulation based calibration.
We demonstrate the capabilities of Octofitter on real and simulated data from different instruments and methods, including HD 91312, simulated JWST/NIRISS aperture masking interferometry observations, radial velocity curves, and grids of images from the Gemini Planet Imager.
We show that Octofitter can reliably recover faint planets in long sequences of images with arbitrary orbital motion.
This publicly available tool will enable the broad application of multi-epoch and multi-method exoplanet detection, which could improve how future targeted ground- and space-based surveys are performed.
Finally, its rapid convergence makes it a useful addition to the existing ecosystem of tools for modelling the orbits of directly imaged planets.

\end{abstract}


\section{Introduction} \label{sec:intro}

Instruments for directly studying exoplanets are steadily improving in sensitivity. 
Current facilities are now accessing planets less than 10 AU from their stars. 
Below these separations, orbital motion can become significant over mere months. 
This will be especially true for facilities with high angular resolving power thanks to their larger apertures and/or shorter operating wavelengths.
This is an advantage for those who wish to determine the orbits of already-known planets but presents a significant challenge when searching for new companions.

When planets move from observation to observation, naive image stacking causes their signals to blur out and fade away.
Reaching a necessary integration in a single epoch hits practical scheduling constraints and eventually physical limitations---for a sufficiently faint planet, it would not be possible to detect a significant number of photons before it moves by a full resolution element.
These constraints apply equally to images as they do to integral field units and interferometers, including aperture masking interferometry (AMI) on JWST \citep{sivaramakrishnanInfraredImagerSlitless2023}, VLTI-GRAVITY \citep{gravitycollaborationFirstLightGRAVITY2017}, and even ALMA \citep{ALMAPaper} in the case of accreting protoplanets.

A number of projects have sought to solve this challenge for image data only by compensating for orbital motion between epochs. These include a search for planets around Sirius B \citep{skemerSiriusImagedMidinfrared2011}, K-Stacker \citep{nowakKStackerKeplerianImage2018,lecorollerKStackerAlgorithmHack2020}, PACOME (Dallant et al., submitted), the search for planets around $\varepsilon$ Eri \citep{mawetDeepExplorationEridani2019,llop-saysonConstrainingOrbitMass2021}, and the search for additional HR 8799 planets by \citet{thompsonDeepOrbitalSearch2023}.
These have now led to promising evidence for $\alpha$ Cen AB b \citep{lecorollerEfficientlyCombiningAlpha2022} and HR 8799 f \citep{thompsonDeepOrbitalSearch2023}.
Moving the analysis of direct imaging data into the orbital domain enables a further extension:
joint models of both images or interferometric observables with indirect exoplanet detection techniques, including radial velocity, astrometric motion, and transit (not directly considered in this paper).
These have previously been explored in \citet{mawetDeepExplorationEridani2019} and \citet{llop-saysonConstrainingOrbitMass2021}.
This opens many possible scenarios. 
In addition to combining images to search for planets despite orbital motion, this allows one to freely combine Doppler or astrometric velocimetry with image data. This can then be used to constrain orbits using images with tentative detections or non-detections, 
improve photometry accuracy or limits,
better constrain a planet's mass,
and/or detect planets where any individual kind of data fails to reach significance \citep[e.g.][]{llop-saysonConstrainingOrbitMass2021}.

These scenarios are possible since all exoplanet detection methods provide orbital constraints that at least partially overlap.
Imaging, RV, and transit all provide the orbital period ($P$); proper motion anomaly (PMA) and RV constrain eccentricity ($e$), the argument of periapsis ($\omega$), and either mass ($m$) or $m \sin(i)$ where $i$ is the orbital inclination;
and multi-epoch imaging/interferometry constrains all orbital parameters up to a $\pm$ ambiguity on the longitude of the ascending node ($\Omega$).
These connections could, in principle, allow information from all methods to flow into a single orbit model and, ultimately, the detection of a new planet.

To apply these ideas broadly, the community will need a tool capable of modelling all different types of exoplanet data. The \texttt{orbitize!} \citep{bluntOrbitizeComprehensiveOrbitfitting2020} and \texttt{orvara} \citep{brandt_orvara_2021_fixed} packages come close:
they support Bayesian modelling of relative astrometry, RV, and PMA.
We needed a publicly available and generally applicable package that goes further to directly model  image and interferometer data with or without independent detections at each epoch.


It is, generally, challenging to accurately compute orbital posteriors since the traditional Campbell orbital elements ($a,e,i,\omega,\Omega,t_\mathrm{peri})$ possess complex co-dependencies and degeneracies (e.g. when $e=0$ or $i=0$).  
This task becomes even more challenging when working with short orbital arcs because they lack the constraining power to independently determine each Campbell element, meaning orbit posteriors are typically complex and very sensitive to their priors \citep[e.g.][]{oneilImprovingOrbitEstimates2019}. 
Introducing image and interferometer data to the model exacerbates issues further, as they produce multi-modal posteriors that are challenging to traverse. 
Any inaccuracies in the calculation of an orbit posterior can lead to errors in mass and/or photometry and a spurious detection.

\begin{figure*}
    \includegraphics[width=\textwidth]{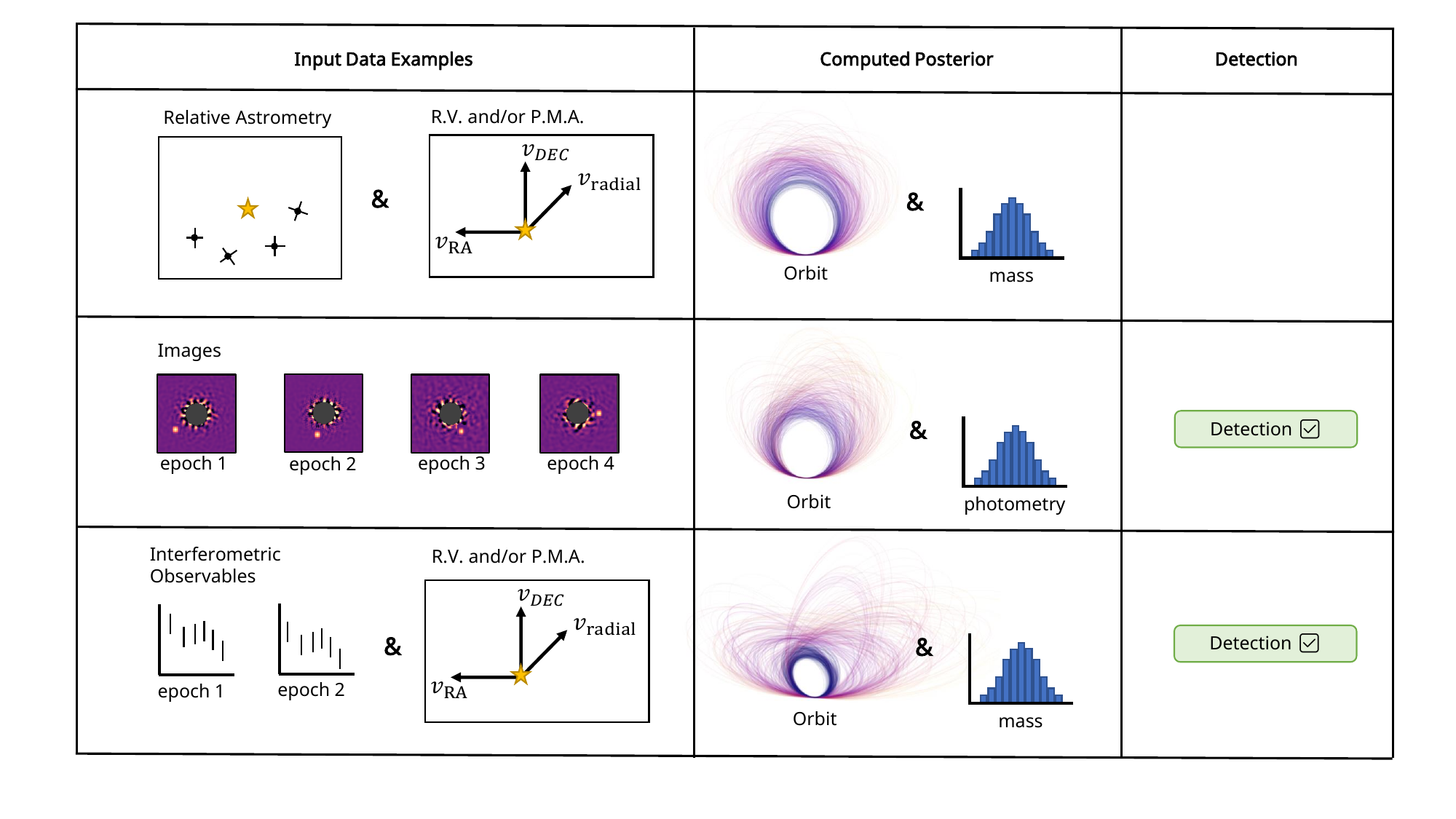}
    \caption{Conceptual schematic of three different ways to use \octo. Other possibilities, like combining images with Doppler or astrometric velocimetry or using a mix of relative astrometry and images without detections to constrain orbits, are also possible.\label{fig:schematic}.}
\end{figure*}

These challenges motivate the development of our new orbit- and data-modelling framework, ``Octofitter.''\footnote{\octo\ is publicly available on the Julia General registry, and extensive usage examples are provided at \url{https://sefffal.github.io/Octofitter.jl/dev/}}
Named for the eight types of data through which it aims to grasp new planets, \octo\ is designed from the ground up to be a flexible platform for modelling and experimentation while providing very high computational performance.
These advances are thanks to our implementation of a pure-Julia \citep{bezansonJuliaFastDynamic2012}, fully differentiable, and non-allocating modelling language, and our use of the higher-order No-U-Turn Hamiltonian Monte Carlo sampler \citep{xuAdvancedHMCJlRobust2020}.
We further present the use of Simulation-Based Calibration \citep[SBC;][]{taltsValidatingBayesianInference2020} to confirm the accuracy of orbital posteriors, which is made practical by \octo's speed. 
Figure \ref{fig:schematic} shows a schematic of a few of the ways \octo\ can be applied to exoplanet data.

As a publicly available code with these advances, \octo\ will enable the wide application of multi-epoch, multi-instrument, and multi-method exoplanet detection and modelling.
These approaches have the ability to improve how direct surveys are completed and improve the yield of the upcoming Nancy Grace Roman Space Telescope Coronagraphic Instrument \citep{kasdinNancyGraceRoman2020}, the Habitable Worlds Observatory (HWO), and future facilities for extremely large telescopes.

\section{Data Models}

    This section describes how the eight kinds of exoplanet data can be modelled by \octo.
    These types of data are relative astrometry,  images, interferometric observables,  star and planet radial velocity (absolute and relative), and proper motion anomaly.

    The \octo\ framework is structured around three concepts:
    likelihood functions for observations,
    system models to tie observations to parameters,
    and generative functions to create synthetic observations. 
    These break the problem down into orthogonal components that can be freely combined to solve a wide range of orbit modelling problems.



    
    Each kind of observation in \octo~is supported by its own \julia\ data type.
    Every observation type is a wrapper for a data table of observations with a preset list of required columns.
    The observational data, in turn, can be provided to its associated type directly in the code, loaded from a local CSV or Arrow file, or obtained from a remote SQL database.
    For each observation type, a method of the \verb|Octofitter.ln_like| function is provided that computes the likelihood of the data it contains given a specific set of parameters. 
    We now describe the observation types and likelihood functions included in \octo.

    \subsection{Relative Astrometry}

    The position measured between a directly imaged planet and its host star is one of the most fundamental measurements gathered from direct observations. 
    It can be extracted from any image where the planet is robustly detected in a single epoch.
    Relative astrometry can be expressed either as separation in milliarcseconds and position angle in degrees,
    or as offsets in milliarcseconds in the R.A.--Dec. tangent plane.
    
    Relative astrometry measurements can be provided to \octo~using \texttt{AstrometryLikelihood}, which accepts a table with columns for
    \texttt{epoch}, the date of the measurement in units of modified Julian days; 
    \texttt{sep}, the separation between primary and secondary in mas;
    \texttt{pa}, the position angle of the secondary measured East from North;
    \texttt{σ\_pa}, \texttt{σ\_sep}, and \texttt{cor} to specify the measurement uncertainties and correlation, respectively.
    Alternatively, \texttt{ra} (relative), \texttt{dec} (relative), \texttt{σ\_ra}, and \texttt{σ\_dec} can be substituted if preferred. 
    A sample relative astrometry input table is presented in Table \ref{tab:rel-astrom}.
    \begin{deluxetable}{cccccc}
        \tablecaption{Relative astrometry input sample. \label{tab:rel-astrom}}
        
        \tablehead{
            \colhead{\texttt{epoch}} & \colhead{\texttt{pa}} &  \colhead{\texttt{sep}} &  \colhead{\texttt{σ\_sep}} & \colhead{\texttt{σ\_pa}} & \colhead{\texttt{cor}} \\
            \colhead{(mjd)} & \colhead{($\degr$)} &  \colhead{(mas)} &  \colhead{(mas)} & \colhead{($\degr$)} & \colhead{}
        }
        \startdata
        58849.0 &  224.93	&   615.2   &   30.0	&   0.8 &	0.0 \\
        58879.0 &  228.53	&   606.4   &   30.0	&   0.8 &	0.0 \\
        58909.0 &  229.66	&   663.3   &   30.0	&   0.8 &	0.0 \\
        58939.0 &  232.48	&   635.9   &   30.0	&   0.8 &	0.0 \\
        58969.0 &  233.25	&   610.3   &   30.0	&   0.8 &	0.0 \\
        58999.0 &  235.22	&   669.7   &   30.0	&   0.8 &	0.0 \\
        59029.0 &  236.67	&   666.8   &   30.0	&   0.8 &	0.0 \\
        59059.0 &  237.85	&   654.3   &   30.0	&   0.8 &	0.0 \\
        59089.0 &  239.68	&   713.6   &   30.0	&   0.8 &	0.0 \\
        59215.0 &  245.67	&   747.3   &   30.0	&   0.8 &	0.0 \\
        59245.0 &  245.54	&   736.4   &   30.0	&   0.8 &	0.0 \\
        59275.0 &  247.94	&   709.7   &   30.0	&   0.8 &	0.0 \\
        59305.0 &  248.13	&   791.4   &   30.0	&   0.8 &	0.0 \\
        59335.0 &  251.17	&   777.8   &   30.0	&   0.8 &	0.0 \\
        59365.0 &  251.43	&   773.8   &   30.0	&   0.8 &	0.0 \\
        59395.0 &  250.44	&   866.5   &   30.0	&   0.8 &	0.0 \\
        59425.0 &  253.49	&   789.5   &   30.0	&   0.8 &	0.0 \\
        59455.0 &  253.98	&   839.1   &   30.0	&   0.8 &	0.0 \\
        59945.0 &  268.72	&   986.3   &   30.0	&   0.8 &	0.0 \\
        59975.0 &  268.89	&   941.6   &   30.0	&   0.8 &	0.0 \\
        60005.0 &  269.96	&   959.2   &   30.0	&   0.8 &	0.0 \\
        60035.0 &  270.0	&   928.9   &   30.0	&   0.8 &	0.0 \\
        60065.0 &  270.67	&   952.9   &   30.0	&   0.8 &	0.0 \\
        60095.0 &  272.02	&   977.5   &   30.0	&   0.8 &	0.0 \\
        60125.0 &  270.99	&   950.1   &   30.0	&   0.8 &	0.0 \\
        60155.0 &  272.15	&   953.9   &   30.0	&   0.8 &	0.0 \\
        60185.0 &  274.33	&   985.2   &   30.0	&   0.8 &	0.0 \\
        \enddata
    \end{deluxetable}
    
    Based on these quantities, the likelihood function is simply
    taken as the Gaussian likelihood that the residual between the model and measurement for each parameter would be seen, given the provided measurement uncertainty. For relative positions measured along the R.A. and Dec. axes, this is
    \begin{equation}\label{eq:lnlike-relastrom}
        \begin{split}
        \log{\mathcal{L}_{\mathrm{astrom}}} = 
            &-\frac{1}{2} \frac{(\Delta \mathrm{RA} - \mathrm{\Delta \tilde{RA}})^2}{\sigma^2_\mathrm{\tilde{\Delta RA}} }
            -\log{\sqrt{2\pi\sigma^2_\mathrm{\tilde{\Delta RA}}}}\\
            &-\frac{1}{2} \frac{(\mathrm{\Delta DEC} - \mathrm{\tilde{DEC}})^2}{\sigma^2_\mathrm{\tilde{\Delta DEC}} }
            - \log{\sqrt{2\pi\sigma^2_\mathrm{\tilde{\Delta DEC}}}},
        \end{split}
    \end{equation}
    where $\mathrm{\Delta RA}$ and $\mathrm{\Delta DEC}$ are the separation from the star along the R.A. and Dec. axes, $\sigma_\mathrm{\Delta RA}$ and $\sigma_\mathrm{DEC}$ are the corresponding uncertainties,
    and where the tilde distinguishes measured quantities from  values calculated from model parameters.
    A more complex expression is used when the correlations between uncertainties are non-zero.

    \subsection{Images}

    Directly modelling point sources in images is one of the key features of \octo.
    This is accomplished with the same approach described in \citet{ruffioBayesianFrameworkExoplanet2018} and applied in \citet{mawetDeepExplorationEridani2019}, \citet{llop-saysonConstrainingOrbitMass2021}, and \citet{thompsonDeepOrbitalSearch2023}.
    In the case where a planet is robustly detected in each image and the uncertainties are well-approximated by Gaussian uncertainties, it is mathematically equivalent to extracting relative astrometry and photometry and then modelling those measurements.
    Where this approach goes beyond the two-step process, however, is when a planet is not detectable in a single epoch or when one wants to constrain the orbit of a planet based on a non-detection.
    The first case may arise when searching for planets that are too faint to detect before they exhibit orbital motion. The second case may arise any time a monitored planet passes too close to its star to detect. In this case, it's not possible to extract astrometry. By directly modelling the images, large swathes of the orbital parameter space may still be ruled out. This can improve constraints on orbital parameters and improve predictions of the planet's future location.

    In \octo, image data can be modelled across any number of photometric bands, instruments, and epochs. Data can be provided using the \texttt{ImageLikelihood} observation type which accepts a data table with the columns \texttt{epoch}, \texttt{band}, \texttt{platescale}, \texttt{image}, and \texttt{contrast}.
    A sample input table is presented in Table \ref{tab:imgs}.
    The platescale and parallax system parameters (which may be fixed or fit to the data) are used to map an orbit to a pixel location in the image. The \texttt{contrast} entry for each image allows one to pass an arbitrary function of position that gives $1\sigma$ contrast. If not provided, \octo~calculates a contrast curve automatically from the image itself. This is sufficient for images with no clear detection, but should be avoided if the image contains a bright planet as the contrast curve will overestimate the noise at the planet's separation.
    Extended emission from disks is not currently considered.

    Given this data, the likelihood function used by \octo\ for each image by \octo~is that of \citet{ruffioBayesianFrameworkExoplanet2018}:

    \begin{equation} \label{eq:lnlike-img}
        \log{\mathcal{L}_\mathrm{img}} = 
            \frac{1}{2\sigma_{B,x,i}^2}  (f_B^2 - 2f_B \tilde{f}_{B,x,i}),
    \end{equation}
    where $f_B$ is the model flux parameter for the photometric band $B$, $x$ is the position in the image determined from  orbital parameters, $i$ is the epoch,  $\sigma$ is the uncertainty,
    and $\tilde{f}_B$ is the measured flux at that same location.

    This likelihood function assumes that flux is constant from epoch to epoch, but could easily be adapted to use an orbital phase function for planets imaged in reflected light \citep[e.g.][]{pogorelyukDeconfusingDetectionsDirectly2022}. Taken to the extreme, the flux can be fit independently at each epoch \citep[e.g.][]{nowakKStackerKeplerianImage2018} at the expense of reduced constraining power.
    For both images and relative astrometry, we do not currently consider uncertainty in the instrument's North-angle or platescale. These might need to be considered when combining data from multiple instruments, in which case they could be added straightforwardly.
    We also note that this likelihood function assumes Gaussian distributed noise in each image. Where this is not a good assumption, standard techniques for inflating uncertainties could be used before applying \octo.
    Furthermore, spatial correlations between pixels are not currently handled.
    Finally, this likelihood function is only specified up to a constant and is not suitable for techniques like nested sampling.

    \begin{deluxetable}{ccccc}
        \tablecaption{Images input sample\label{tab:imgs}}
        
        \tablehead{
            \colhead{\texttt{epoch}} & \colhead{\texttt{band}} &  \colhead{\texttt{platescale}} &  \colhead{\texttt{image}} &  \colhead{\texttt{contrast}} \\
            \colhead{(mjd)} & \colhead{(symbol)} & \colhead{(mas/px)} &  \colhead{(matrix)} & \colhead{(function)}
        }
        \startdata
        59976.0 & \texttt{:H} & 10.2 & \dots & \dots \\
        59976.0 & \texttt{:J} & 10.2 & \dots & \dots \\
        60576.0 & \texttt{:H} & 10.2 & \dots & \dots \\
        \enddata
    \end{deluxetable}

    \subsection{Interferometric Observables}

    Just as with imaging, combining interferometric observations across multiple epochs using orbital modelling removes the need to detect a companion in a single observation.
    We implement a model assuming an unresolved point source primary and $N$ unresolved point source companions. The complex visibilities of this model are given by
 
    \begin{equation} \label{eq:vis-bin}
        \mathrm{V}_{\mathrm{bin}} = \frac{1+ \sum_{i=1}^{N} f_i\mathrm{exp}\Big(-2 \pi \mathrm{i} \Big(\Delta \mathrm{RA}_i u + \Delta \mathrm{DEC}_i v \Big) \Big)}{1+ \sum_{i=1}^{N} f_i},
    \end{equation}
    where $f_i$ is the (companion/primary) contrast of the $i$th companion, $\Delta$RA$_i$ and $\Delta$DEC$_i$ are the right ascension and declination of the $i$th companion, and $u$ and $v$ are the Fourier domain coordinates, with magnitudes given by the interferometer baseline lengths divided by the observing wavelength \citep[e.g.,][]{2003EAS.....6...23B, kammererInfraredImagerSlitless2023}.
    A sample input table is presented in Table \ref{tab:interf}.

    The squared visibilities are calculated from the squared moduli of the complex visibilities, and the closure phases are calculated by summing the phases of the three complex visibilities calculated from triangles of stations in the interferometer. We construct likelihood functions for the squared visibilities and the closure phases separately, assuming Gaussian noise statistics and diagonal covariance matrices. We also assume that there is at least a moderate contrast between the primary and any companions such that we can neglect phase wrapping in the closure phase \citep{Thiebaut:17}. These likelihood functions are given by

    \begin{equation} \label{eq:lnlike-cp}
        \log{\mathcal{L}_{\mathrm{CP}}} = -\frac{1}{2}\sum_i \frac{(\mathrm{CP}_i - \tilde{\mathrm{CP}_i})^2}{\sigma_{\mathrm{CP},i}^2} -\frac{1}{2} \log{2\pi \sigma_{\mathrm{CP},i}^2}
    \end{equation}  
    and
    \begin{equation} \label{eq:lnlike-v2}
        \log{\mathcal{L}_{\mathrm{V^2}}} = -\frac{1}{2}\sum_i \frac{(\mathrm{V^2}_i - \tilde{\mathrm{V^2}_i})^2}{\sigma_{\mathrm{V^2},i}^2} -\frac{1}{2} \log{2\pi \sigma_{\mathrm{V^2},i}^2},
    \end{equation}
    where CP$_i$ is the $i$th closure phase, V$^2_i$ is the $i$th squared visibility and $\sigma_{\mathrm{CP},i}$, and $\sigma_{\mathrm{V}^2,i}$ are the uncertainties in the $i$th squared visibility and closure phase, respectively.
       

    \begin{deluxetable}{ccccc}
        \tablecaption{Visibilities input sample\label{tab:interf}}
        
        \tablehead{
            \colhead{\texttt{epoch}} & \colhead{\texttt{band}} &  \colhead{\texttt{pa}} &  \colhead{\texttt{sep}} &  \colhead{\texttt{contrast}} \\
            \colhead{(mjd)} & \colhead{(symbol)} & \colhead{($\degr$)} &  \colhead{(mas)} & \colhead{}
        }
        \startdata
        60096.0 & \texttt{:F480M} & -92.7 & 180.5 & 0.00036 \\
        60171.0 & \texttt{:F480M} & -61.5 & 159.1 & 0.00036 \\
        60462.0 & \texttt{:F480M} & 56.9 & 213.9 & 0.00036 \\
        \enddata
    \end{deluxetable}

    \subsection{Radial Velocity}

    Similarly to other packages, \octo~allows one to model radial velocity data in combination with other observation types.
    If combined with relative astrometry, proper motion anomaly, or image data, this allows one to directly model the dynamical mass of a planet.
    
    RV data can be specified for the star or the planet using \texttt{RadialVelocityLikelihood}, which accepts a table with columns for \texttt{epoch} in MJD,
    \texttt{rv} in m/s, 
    \verb|σ_rv|, the uncertainty on \texttt{rv} in the same units,
    and \verb|inst_idx|.
    \verb|inst_idx| is an integer between 1 and 4 used to specify which instrument the measurement corresponds to.
    The zero point \verb|rv0_i| and jitter \verb|jitter_i| must be specified as variables in the model where \texttt{i} corresponds to the instrument index.
    
    A sample input table is presented in Table \ref{tab:rv}. The input format is the same for both cases. Depending on how the zero point is modelled, it is possible to use either relative or absolute RVs. The zero point can also modelled as an arbitrary function of other variables, allowing one to fit, for example, linear trends.

    When combining radial velocity data with direct imaging modelling, it is possible (though not required) to connect the planet's photometry with its dynamical mass by using a user-supplied model.
    This model can either map a mass and/or system age to photometry in each band, or vice-versa. Connecting these two variables may be useful in cases such as when the orbit is determined by radial velocity up to the inclination degeneracy but has not yet been detected with direct imaging. 
    
    
    In a similar manner to relative astrometry, we define a radial velocity likelihood function that allows us to fit orbital parameters to radial velocity data. This function is
    \begin{equation} \label{eq:lnlike-rv}
            \log{\mathcal{L}_{\mathrm{rv}}} = -\frac{1}{2}\sum_i \frac{(v_{\text{r},i} - \tilde{v}_{\text{r},i})^2}{\sigma_{v_{\text{r}},i}^2} - \log{2\pi \sigma_{v_{\text{r}},i}^2},
    \end{equation}
    where $v_{\text{r},i}$ is the measured radial velocity, $\tilde{v}_{\text{r},i}$ is the maximum likelihood estimate of the radial velocity, and $\sigma_{v_{\text{r}},i}$ is the uncertainty in the radial velocity, all at the epoch $i$.

    \octo~supports multiple instruments, each with their own RV zero-point and jitter term and all with independently selectable priors. This is accomplished via the \texttt{inst\_idx} column which associates each RV measurement to a particular instrument, zero point, and jitter.
    As a convenience, \octo~includes a helper function to load RV curves from the publicly available HARPS RV Bank \citep{trifonovPublicHARPSRadial2020}.
    Accessing this data will prompt the user to accept its license and will automatically fetch the RV curves.
  

    \begin{deluxetable}{cccc}
        \tablecaption{Radial velocity input sample\label{tab:rv}}
        
        \tablehead{
            \colhead{\texttt{epoch}} & \colhead{\texttt{rv}} &  \colhead{\texttt{σ\_rv}} &  \colhead{\texttt{inst\_idx}} \\
            \colhead{(mjd)} & \colhead{(m/s)} & \colhead{(m/s)} & \colhead{}
        }
        \startdata
        58849.0 &   47.2342   & 5.0 & 1        \\
        58879.0 &   14.4347   & 5.0 & 1        \\
        58909.0 &   12.9645   & 5.0 & 1        \\
        58939.0 &   26.2633   & 5.0 & 1        \\
        58969.0 &   -8.27905   & 5.0 & 1        \\
        58999.0 &   4.94685   & 5.0 & 1        \\
        59029.0 &   0.863664   & 5.0 & 1        \\
        59059.0 &   10.9524   & 5.0 & 1        \\
        59089.0 &   3.92389   & 5.0 & 1        \\
        59215.0 &   21.0509   & 5.0 & 1        \\
        59245.0 &   29.2009   & 5.0 & 1        \\
        59275.0 &   16.77   & 5.0 & 1        \\
        59305.0 &   19.0421   & 5.0 & 1        \\
        59335.0 &   52.9403   & 5.0 & 2       \\
        59365.0 &   21.8173   & 5.0 & 2       \\
        59395.0 &   55.0851   & 5.0 & 2       \\
        59425.0 &   20.444   & 5.0 & 2       \\
        59455.0 &   0.915145   & 5.0 & 2       \\
        59945.0 &   21.6418   & 5.0 & 2       \\
        59975.0 &   -0.54618   & 5.0 & 2       \\
        60005.0 &   11.4391   & 5.0 & 2       \\
        60035.0 &   15.3402   & 5.0 & 2       \\
        60065.0 &   17.7835   & 5.0 & 2       \\
        60095.0 &   5.35458   & 5.0 & 2       \\
        60125.0 &   60.3179   & 5.0 & 2       \\
        60155.0 &   43.5441   & 5.0 & 2       \\
        60185.0 &   11.4066   & 5.0 & 2       \\
        \enddata
    \end{deluxetable}

    \subsection{Proper Motion Anomaly}
    
    Many systems that are candidates for direct imaging have their positions and proper motions measured accurately by both the Gaia \citep{gaia-collaborationGaiaEarlyData2021} and Hipparcos \citep{vanleeuwenValidationNewHipparcos2007} missions. The Hipparcos-Gaia Catalog of Accelerations \citep[HGCA][]{brandt_hipparcos-gaia_2021_fixed} cross-calibrates measurements from these two satellites and inflates uncertainties to cover most known systematics.
    This results in projected velocity measurements of the system's photocentre at the Hipparcos epoch, the Gaia epoch, and between the two.

    In \octo, proper motion anomaly must be loaded directly from the HGCA by specifying the host star's Gaia identifier (the DR3 version at the time of writing). As with the radial velocity loaders, this will prompt the user to accept a license for the catalog and automatically download the HGCA.
    
    In order to connect proper motion anomaly with the orbital image modelling described above, we define a likelihood function based on the HGCA data and the same orbital parameters:
    
    \begin{equation} \label{eq:lnlike-pma}
        \begin{split}
            \log{\mathcal{L}_{\mathrm{pma}}} =
            &-\frac{1}{2} 
                \frac{\left (v_{\mathrm{RA}} -  \tilde{v}_{\mathrm{RA}} \right)^2}{\sigma^2_{\tilde{v}_{\mathrm{RA}}}} - \log{2\pi\sigma^2_{\tilde{v}_{\mathrm{ RA}}}} \\
            &-\frac{1}{2} 
                \frac{(v_{\mathrm{DEC}} - \tilde{v}_{\mathrm{DEC}})^2}{\sigma^2_{\tilde{v}_{\mathrm{DEC}}}} - \log{2\pi\sigma^2_{\tilde{v}_{\mathrm{ DEC}}}},
        \end{split}
    \end{equation}
    \noindent where $v_{\mathrm{ RA}}$ and $\tilde{v}_{\mathrm{DEC}}$ are the R.A. and Dec. proper motion at each epoch and $\sigma_{\mathrm{RA}}$ and $\sigma_{\tilde{v}_{\mathrm{ DEC}}}$ are the associated uncertainties.
    The input table format for this observation type matches Table 4 of  \citet{brandt_hipparcos-gaia_2021_fixed}.
    
    Compared with \citet{brandt_orvara_2021_fixed} and \citet{brandt_htof_2021_fixed}  we adopt a simpler model of the Gaia and Hipparcos data. Rather than use the exact epochs each mission scanned the star (or estimate using the Gaia Observation Forecast Tool\footnote{\url{https://gaia.esac.esa.int/gost/}}), we presume that the position and proper motion of the star were measured independently 25 times over the duration of each mission. For orbits with periods longer than the observation windows of Gaia ($\approx 668$ days) and Hipparcos ($\approx 1227$ days), using these 25 epochs approximates the smearing effect caused by the moving photocenter during observations.
    Future work could refine this model further following their more sophisticated approach by re-implementing the calculations of HTOF \citep{brandt_htof_2021_fixed} in Julia.
    Once it is available, it should also be possible to use the full intermediate GAIA astrometry catalogue in Octofitter.

\section{Methods}
    
\subsection{Assessing Detections}
This section describes how we choose to interpret the results of a model fit with \octo, though it should be noted that the modelling code makes no such prescriptions. 
For evaluating upper limits and detections from image or visibility data alone, we follow the conventions of \citet{ruffioBayesianFrameworkExoplanet2018}.


As set out in \citet{ruffioBayesianFrameworkExoplanet2018}, we calculate upper photometry or mass limits by finding where the cumulative distribution function equals some threshold e.g. 97.7\%. That is, $\mathrm{CDF}\left(f_\mathrm{lim}\right) = 97.7\%$.
This can be stated as an overall value for the system to say, for example, that we believe with 97.7\% confidence that there are no planets present at all with photometry above some $f_\mathrm{lim}$. 
It can also be calculated over small ranges of orbital parameters, in order to, for example, calculate a flux upper limit as a function of semi-major axis.
This value will be driven by the brightest speckles close to the star, so a much more useful metric is to provide $f_\mathrm{lim}$ as a function of orbital parameters.

This model does presuppose that a planet exists and follows a Keplerian orbit around the star; however, as long as the photometry prior is broad, the model is free to drive the photometry towards zero.
We can therefore use the photometry posterior to calculate a signal to noise ratio (SNR) analogous to the typical metric used in direct imaging.
This approach has the benefit of being familiar to those with experience evaluating direct images for detections.

\begin{figure}
    \centering
    \includegraphics[width=\colwidth]{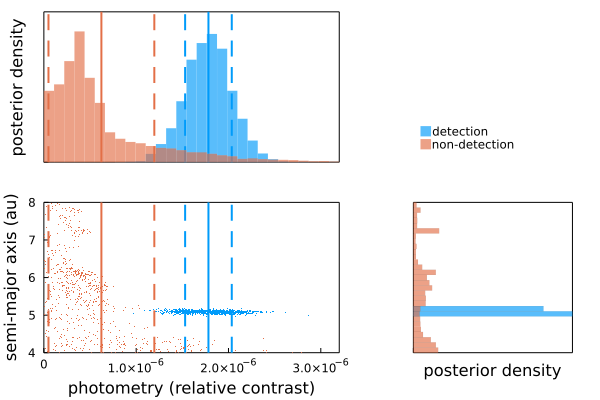}
    \caption{
        Posterior density of photometry versus semi-major axis for a simulated detection and non-detection.
        The solid lines mark the mean of the marginal photometry posterior, and the dashed lines mark $\pm 1\sigma$.
        We consider a planet detected when the SNR based on the photometry marginalized over all other parameters is greater than some chosen threshold, e.g. $5\sigma$.
    }
    \label{fig:detection-example}
\end{figure}

The procedure we adopt for assessing detections is to
\begin{enumerate}
    \item sample from the posterior,
    \item marginalize over all orbital parameters,
    \item calculate the SNR as the mean of the marginal photometry posterior divided by its standard deviation,
    \item and then compare this SNR with a pre-selected threshold based on a tolerable false positive rate, e.g. $5\sigma$.
\end{enumerate}
The results of this procedure are visualized in Figure \ref{fig:detection-example}.
This gives a single SNR value for existence of any planet in that dataset consistent with the user's choice of priors.
This is somewhat different than the typical SNR detection thresholds used because this SNR is calculated by marginalizing over all credible orbits, or equivalently, if applied to a single image, marginalizing over all credible positions.
By contrast, the standard $5\sigma$ threshold is applied to the position in an image with the highest SNR, rather than a weighted combination of all positions.

In general, SNRs obtained using this approach assume that
\begin{enumerate}
    \item the noise is approximately Gaussian distributed,
    \item the noise is uncorrelated along orbital tracks between images,
    \item the images contain only a single planet,
    \item and that planet closely follows a Keplerian orbit.
\end{enumerate}
Deviations from assumptions (1-2) may increase the false positive rate, while deviations from (3-4) may lower the recovered photometry and SNR.

Multi-planet systems could be accommodated easily by introducing additional planets to the model and/or restricting the priors to include only a subset of the parameter space.


For models that combine direct and indirect data, the relationship between mass and photometry variables may be complex, and this simple scheme may not be appropriate. In this case, it may be better to calculate a Bayes factor between planet and no-planet models.
Such a Bayes factor can be treated as a direct measurement of our belief that a planet exists and follows a Keplerian orbit.
This can be carried out in a limited fashion in \octo\ by calculating the Savage-Dickey density ratio of the mass or photometry marginal posterior (\cite{dickeyWeightedLikelihoodRatio1971,koopBayesianEconometrics2003}; for a pedagogical text, see \citet{wagenmakersBayesianHypothesisTesting2010}).
This is more flexible since it does not require a uniform prior---any prior that includes zero mass or photometry would suffice---and because it does not presume that the marginal posterior is Gaussian distributed.

    \subsection{Orbital Bases and Priors}

    \octo\ supports a wide range of different orbital bases for use in different situations. These include traditional Campbell elements ($a,e,i,\omega,\Omega$), Thiele-Innes elements \citep[$e,A,B,F,G$,][]{hartkopfBinaryStarOrbits1989,wrightEfficientFittingMultiplanet2009,oneilImprovingOrbitEstimates2019}, Cartesian elements \citep[$x,y,z,v_x,v_y,v_z$,][]{ferrer-chavezAlgorithmicSpeedupsPosterior2021}, and a reduced basis set for modelling radial velocity only ($a, e, m \sin(i), \omega$).
    Users can specify priors using arbitrary distributions from \texttt{Distributions.jl}\footnote{\url{https://juliastats.org/Distributions.jl}} and functions of those distributions.

    For the analyses presented in this paper, we adopt either Campbell elements or Thiele-Innes elements with the following priors and modifications.
    When using Campbell elements, we adopt a log-uniform prior on semi-major axis, a uniform prior on eccentricity, a sine prior on inclination, a Gaussian prior on host mass, and uniform priors on the remaining angular parameters.
    Note \octo\ reports the argument of periastron of the planet (and not the star) as $\omega$ and adopts $+z$ increasing away from the observer. These conventions match those used by \texttt{orvara} and \texttt{orbitize!} \citep{householder2023inconsistent}. The full conventions used in \octo\ are described in Appendix \ref{sec:derivation}.
    When using Thiele-Innes elements, we adopt a log uniform prior on a ``scale'' parameter and multiply this with standard Gaussian priors on the constants A, B, F, and G. This maintains a log-uniform prior on semi-major axis and sine prior on inclination.
    For both cases, we replace the parameter $\tau$, which gives the position of a planet 
    along its orbit
    \citep{bluntOrbitizeComprehensiveOrbitfitting2020} 
    , with $\theta$, the position angle at the average epoch of the observations.
    This improves convergence relative to sampling from the $\tau$ parameter directly since $\theta$ is directly constrained by relative astrometry and is not sensitive to other orbital parameters.
    A derivation of $\tau$ from position angle $\theta$ is presented in Appendix \ref{sec:theta-tau-derivation}.

    \subsection{Modelling Language}
    
    To specify the structure of a system model, \octo~provides a domain-specific modelling language. This simple language allows for the parameterizations and observations associated with each planet and the host star to be independently specified. For example, we could attach separate radial velocity measurements to each object in a system.
    Orbital parameters can be drawn from arbitrary prior distributions, fixed to particular values, or computed from combinations of other system parameters. 

    This modelling language makes it convenient to work with simple systems, like fitting the orbit of one planet to relative astrometry measurements, as well as more complex multi-planet systems.

    The following is an example of a planet model using traditional Campbell orbital parameters, user-specified priors, and relative astrometry data:
    \begin{samepage}
    \begin{lstlisting}[language = Julia]
table = Table(CSV.read("astrom.csv"))
astrom = AstrometryLikelihood(table)
@planet b Visual{KepOrbit} begin
    a ~ LogUniform(2.5, 25)
    i ~ Sine()
    e ~ truncated(
            Normal(0.2, 0.2),
            lower=0,
            upper=1
    )
    Ω ~ UniformCircular()
    ω ~ UniformCircular()
    τ ~ UniformCircular(1.0)
end astrom
    \end{lstlisting}
    \end{samepage}

    A similar syntax is used to specify stellar properties:
    \begin{samepage}
    \begin{lstlisting}[language = Julia]
@system HD1234 begin
    plx ~ truncated(
            Normal(41.123, 0.012), lower=0)
    M ~ truncated(
            Normal(1.5, 0.05), lower=0)
    age_Myr ~ Uniform(30, 300)
end b
    \end{lstlisting}
    \end{samepage}

    The planet and system definition blocks contain pairs of variable names and values
    which can be constants,
    prior distributions,
    or arbitrary functions of other variables.
    Variables drawn from priors are specified by \verb|~| whereas 
    variables defined as constants or calculated as a function of other variables are defined with \verb|=|.
    In this example,
    the convenience function \texttt{UniformCircular} creates two independent variables with standard normal priors
    and computes the angle between them using the arctangent.
    This, in effect, creates a circular domain where samples can smoothly wrap past 0 and $2\pi$. 
    Users are free to create their own parameterizations and likelihoods and combine them arbitrarily. 
    The only requirements are that they are differentiable, smooth, and return a finite value at all points in the domain given by the priors.

    This modelling approach, by being declarative rather than imperative, as in \texttt{exoplanet.py} \citep{Foreman-Mackey_exoplanet_Gradient-based_probabilistic_2021},
    allows us to transform and evaluate the model in several ways. 
    One key restriction is that each prior is proper,
    meaning it is a true probability distribution that integrates to unity.
    This is in contrast to some of the defaults used by \texttt{orvara}, which adopts fully scale-independent,
    but improper (in the statistical sense of the word) log-uniform priors.
    \octo~requires proper priors to support tasks that require sampling directly from the prior distributions, such as simulation-based calibration,
    as will be discussed in Section \ref{sec:sbc}.

    \subsection{Numerical Methods}

    From a model definition, \octo~can generate efficient machine code using runtime-generated functions and Julia's just-in-time compiler.
    This code generation step allows \octo\ to support a rich variety of observation types without paying any runtime overhead for features that aren't used.
    
    For the purposes of sampling from the posterior,
    \octo~begins by remapping all variables from their possibly limited support (for example, eccentricity constrained between 0 and 1) 
    into new variables defined across the entire real line.
    This makes it so that by construction,
    invalid parameter values like negative masses or semi-major axes are not possible to construct and will not be hit by a sampler. 
    This transformation is performed by the Julia package \texttt{Bijectors.jl}\footnote{\url{https://turinglang.org/Bijectors.jl/}}.

    Next,
    a log-prior function is created for the model that simply evaluates the log-probability density of each prior distribution given a set of parameters.
    To preserve the user-specified prior distributions in place of the automatic bijections,
    a correction is applied based on the Jacobian of the transformation.

    Similarly,
    a log-likelihood function is created based on the provided model and observations.
    Various constants are pre-calculated and reused between orbit solutions.

    To enable the use of higher-order samplers, including Hamiltonian Monte Carlo,
    forward mode automatic differentiation is used to differentiate through the generated log-prior and log-likelihood functions \citep{revelsForwardModeAutomaticDifferentiation2016}.
    This provides the gradient of the log-posterior density, in addition to the value itself, without the overhead of calculating finite differences.
    The overhead of calculating both the log-posterior density and its gradient using forward mode automatic differentiation can be as low as just $2\times$ compared to $10.2\times$ for finite differences.

    Special care was taken to remove all dynamic memory allocations from the generated log-density and gradient functions to prevent overhead from the Julia garbage collector.

    \octo~implements the Julia \texttt{LogDensityProblems} interface so that user models can be sampled from a wide variety of Julia-based MCMC sampler packages, including
    \texttt{AdvancedMH},
    \texttt{AdvancedHMC},
    and \texttt{MCMCTempering}.
    This allows users to select the best sampler for their particular problem and to compare results against samplers used by other popular packages.

    The No-U-Turn Sampler (NUTS) variant of Hamiltonian Monte Carlo \citep{hoffmanNouturnSamplerAdaptively2014} provided by \texttt{AdvancedHMC} is the default used by \octo.
    It allows the code to explore complex posterior distributions with many fewer log-posterior density evaluations by simulating physical trajectories across the posterior landscape.
    The performance of NUTS in \octo~will be discussed in Section \ref{sec:performance}.

    When using the default NUTS sampler,
    we use a dense mass matrix,
    a jittered leapfrog integrator,
    a multinomial trajectory sampler,
    and allow the user to specify a maximum tree depth.
    We initialize the sampler by drawing 250,000 samples from the priors and selecting the value with the highest posterior density as the starting point.
    We also initialize the diagonal elements of the mass matrix using the interquartile range of the priors.
    We found this procedure was more robust than trying to determine the maximum \emph{a posteriori} value with an optimizer for multi-modal posteriors like those found when modelling images.
    In particular, this procedure is less likely than an optimization pass to get stuck in a local optimum or other pathological position.
    A downside of this approach is that it may be inefficient if the priors are narrow and far from the posterior.
    For this case,
    the user can supply a starting point manually as in other tools.
    Next,
    we adapt the mass matrix and step size according to \texttt{AdvancedHMC}'s implementation of the Stan windowed adaptation strategy.
    Finally,
    sampling proceeds until a preset number of accepted proposals are found---typically on the order of 1,000 to 10,000.

    \subsection{Kepler Solver}

    To sample from planet models,
    one must map from the orbital parameters specified in the system model to simulated observations, such as a planet's position over time in the plane of the sky or the radial velocity of the host star.
    In all cases, this requires solving Kepler's equation at every observation epoch and parameter draw.
    Kepler's equation (Eq. \ref{eq:eccentricanomaly}) connects eccentricity and mean anomaly,
    a pseudo-angle somewhat analogous to the amount a planet has moved around its orbit,
    into eccentric anomaly, which can be used to find the actual location of the planet in its orbit.

    Kepler's equation is transcendental and can't be solved analytically outside of special cases. 
    The traditional approach to solving the equation is to use an iterative procedure like Newton's method with an initial guess chosen carefully from the mean anomaly and eccentricity.
    \octo's strategy is to wrap many different pluggable Kepler solvers that can be useful in different scenarios
    and use a fast non-iterative method as a default.
    

A non-exhaustive list of solvers supported by \octo~are
    \texttt{Markley} \citep{markleyKeplerEquationSolver1995},
    \texttt{Goat} \citep{philcoxKeplerGoatHerd2021},
    and \texttt{Newton}.
    Newton's method and many other available root finding algorithms are provided by the \texttt{Roots.jl}\footnote{\url{https://juliamath.github.io/Roots.jl/}} julia package.

    Since these solvers are all implemented in pure Julia, there is no overhead from calling between Python and a C/Cython\footnote{\url{https://cython.org/}} library and full performance is achieved with or without vectorization. 
    Additionally, the solvers support changing the numerical precision between 16- bit, 32-bit, 64-bit, and arbitrary precision. 
    Currently, no particular effort has been made to exploit hardware SIMD vectorization across epochs.

    The \texttt{Markley} algorithm, the default choice, converges to nearly machine precision for all bound orbits when used with 64-bit floating point values. 
    The \texttt{Newton} method can be combined with arbitrary precision floating point numbers to achieve arbitrarily tight tolerances if needed for niche applications.
	The default Markley \citeyear{markleyKeplerEquationSolver1995} based method executes in just 90 ns on 64-bit floating point values for any valid eccentricity and mean anomaly (benchmarked on a Skylake Intel Xeon processor). This is slightly slower than the state-of-the-art results reported by \citet{brandtOrvaraEfficientCode2021}; however, their results of $\approx 60$ ns are only achieved when the solver is vectorized over many epochs. 

    Applying automatic differentiation through an iterative Kepler equation solver would lead to poor performance.
    Even though Kepler's equation has no closed solution, the derivatives of eccentric anomaly with respect to mean anomaly and eccentricity can be found analytically using implicit differentiation (Eqs. \ref{eq:kepler-partial-ma} and \ref{eq:kepler-partial-e}); that is, if we have already solved Kepler's equation for eccentric anomaly, we can calculate its gradient inexpensively.
    We supply these equations as a manual rule to the automatic differentiation library.



    \subsection{Analysis and Visualization}

    Once sampling is complete, \octo~supports the user in testing the convergence of their chain or chains.
    The sampling results are returned as \texttt{Chains} objects from \texttt{MCMCChains.jl}.
    This table-like structure includes entries for each accepted MCMC proposal as well as metadata about the sampling process, such as the compute time used.
    All variables are returned, including those that were fixed to a constant or calculated deterministically from combinations of other variables.
    An automatic summary is output that includes plausible intervals,
    expectation values,
    effective sample sizes (ESS),
    and $\hat{R}$ statistics for each parameter.
    The user can test the convergence of their chain or chains by assessing those ESS and $\hat{R}$ values,
    creating a trace plot,
    and/or by calculating figures like the Gelman, Rubin, and Brooks diagnostic \citep{gelmanInferenceIterativeSimulation1992,brooksGeneralMethodsMonitoring1998} using tools provided by \texttt{MCMCChains.jl} and \texttt{MCMCDiagnosticTools.jl}. 

    \octo~can be used to visualize orbit fits in several ways.
    The orbits represented by chains can be plotted in the plane of the sky,
    in physical system coordinates (i.e. AU),
    or as a time series (e.g. for Doppler or astrometric velocimetry).
    
    \begin{figure*}
        \centering
        \includegraphics[width=\textwidth]{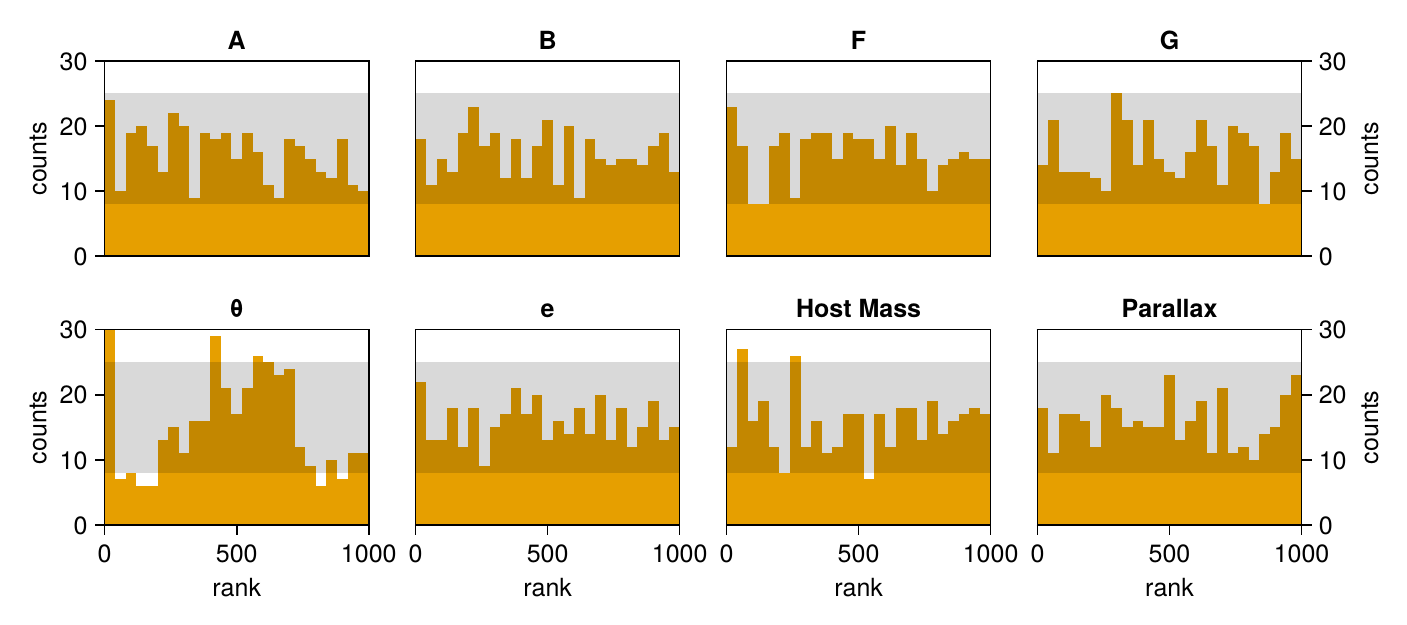}
        \caption{
            Sample results of running the simulation-based calibration procedure on a model consisting of a single planet parameterized with Thiele-Innes elements ($A$, $B$, $F$, $G$, etc.) and position angle $\theta$.
            Each count represents a model fit to a different simulated system.
            The horizontal band gives a 99\% range around the expected value for a perfect sampling procedure.
            The observation epochs and uncertainties are taken from Table \ref{tab:rel-astrom}.
        \label{fig:sbc-grid}}
    \end{figure*}
    
    When visualizing an orbital posterior, a common challenge is ensuring enough data points are used to create a smooth arc.
    This becomes especially challenging with eccentric orbits or ensembles of orbits with widely varying periods.
    Tracing out orbits in equal time steps will lead to most points clustering closer to apoastron, where the planet is moving the most slowly.
    The strategy used by \octo~is to trace orbits in equal steps of eccentric anomaly,
    as suggested in \citet{berryGeneralizedSundmanTransformation2002}.
    This places points in regions of greater curvature, creating smooth arcs with fewer points.


    \octo\ includes functions to generate plots that visualize the posterior and compare it to the input data.
    This includes astrometry, separation, position angle, radial velocity, and proper motion anomaly.
    It also includes a function for visualizing orbits in spatial coordinates (units of AU) in one, two, or three dimensions.
    Examples of these plots are shown throughout the text and documented with the online tutorials.

    \subsection{Simulation-Based Calibration}\label{sec:sbc}
    
    Simulation-based calibration \citep[SBC;][]{cookValidationSoftwareBayesian2006,taltsValidatingBayesianInference2020}
    is a technique that allows one to verify that the output of a Bayesian modelling procedure is unbiased.
    This includes any part of the computation, from model specification to the sampling procedure after the choice of priors.
    Verifying that this choice of priors is reasonable is the domain of other procedures like prior and posterior predictive tests.

    A conceptually related procedure was carried out in \citet{ferrer-chavezBiasesOrbitalFitting2021} in which the authors systematically tested the \texttt{orbitize!} package, parameterization, and default priors for biases. Our contribution here is to present a procedure that is well explored in the statistics literature, is specific to a given set of observation epochs and measurement uncertainties, and can be applied in an automated fashion.
    
    The calibration procedure requires a generative model---that is, a way to take a given set of parameters and create a simulated observation. A familiar example from direct imaging is the injection of fake point sources to test image processing algorithms.
    To follow this procedure, one repeatedly draws a set of parameters
    from the priors, creates simulated observations based on those parameters, samples from the resulting posterior, and then compares the true parameter values (originally sampled from the priors) with the resulting posterior. By doing this many times, one creates a histogram of rank statistics that can reveal many sources of biases present in the model and sampling process. To be clear, this does not evaluate how the choice of priors impacts the posterior.

    
    We implemented support for performing simulation-based calibration automatically in \octo.
    In this procedure, \octo\ takes a given model's observational data, discards the actual measurements, and keeps the epochs and uncertainties associated with each measurement.
    \octo\ then repeatedly creates simulated data at each epoch by drawing from the priors and performs SBC on this simulated data. 
    SBC should, in general, be applied to each model to confirm that it is working as expected.

    \begin{figure*}
        \centering
        \includegraphics[width=0.9\textwidth]{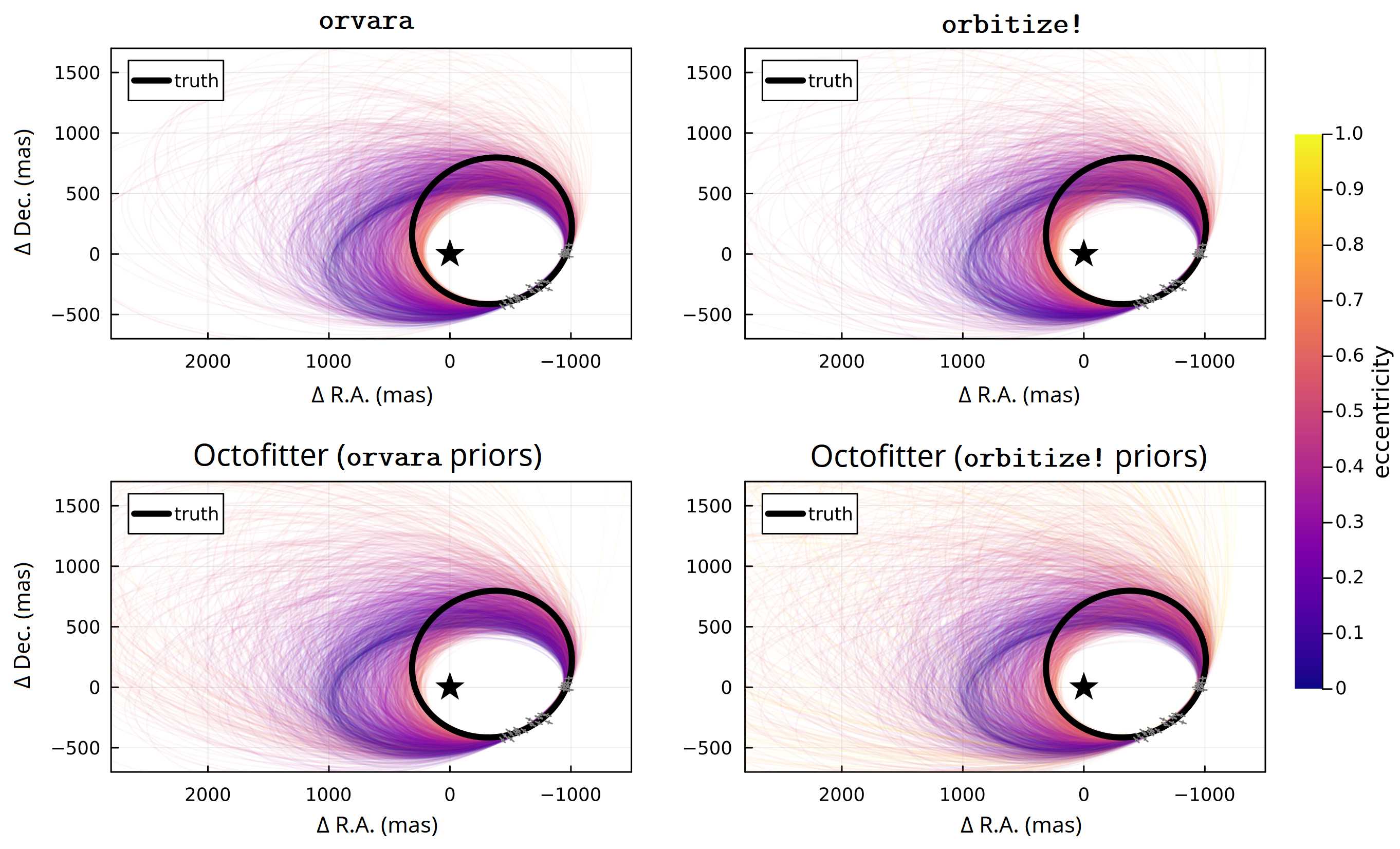}
        \caption{
            Orbit fitting posteriors visualized in the plane of the sky, compared between three packages: \texttt{orvara}, \texttt{orbitize!}, and \octo.
            \label{fig:sampler-comparison}}
    \end{figure*}

    \begin{figure*}
        \includegraphics[width=\textwidth]{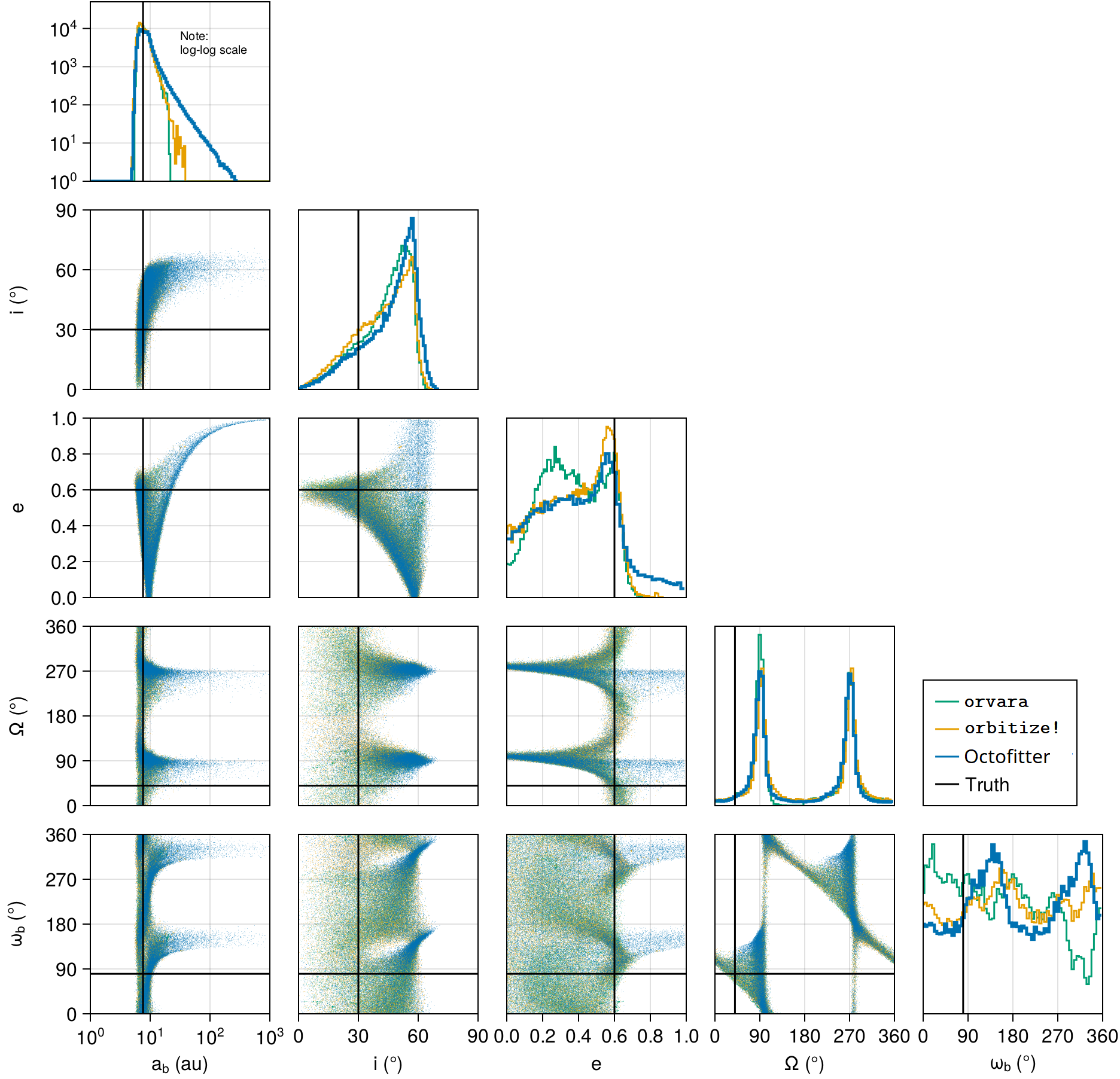}
        \caption{
            Corner plot of key orbital parameters for the case shown in Figure \ref{fig:sampler-comparison} compared between three packages: \texttt{orvara}, \texttt{orbitize!}, and \octo. Semi-major axis is plotted on a log scale to reveal how the sampler behaviour differs in the long tail.
            Though we expect all packages to eventually converge to near-identical results, we find that there are small differences in the $\omega$, $a$, and $e$ marginal posteriors that persist even after running for multiple days.
            \label{fig:sampler-comparison-pairplot}}
    \end{figure*}

    Figure \ref{fig:sbc-grid} shows the results of the simulation-based calibration procedure applied to a model of a planet parameterized by Thiele-Innes elements and position angle at the average epoch.
    For the most part, these histograms do not reveal any systematic biases in the \octo\ sampling procedure.
    One exception is the ``bump'' in the position angle histogram. This shape indicates that \octo\ is under-confident and the sampled posterior distribution is wider than the true posterior.
    By contrast, seemingly small tweaks to the choice of priors can result in histograms with strong biases. 
    For example, drawing the Thiele-Innes constants (A,B) and (F,G) from log-uniform prior distributions rather than Gaussian distributions centred around a single scale, itself drawn from a log-uniform prior, results in noticeable issues with the SBC histograms.
    A guide to interpreting the results of the SBC procedure is available in \citet{taltsValidatingBayesianInference2020}.

    These tests illustrate the value of performing the simulation-based calibration procedure for each new model and data
    combination a user wishes to use.
    Given the complexity of orbit models and the difficulty of sampling from them, we do not expect our sampling procedure to be entirely unbiased.
    Rather, we hope that by creating an easy way to diagnose these biases, users of \octo~can interpret their results with an appropriate level of skepticism in accordance with the level of bias detected.





    \subsection{Other Packages}\label{sec:other-packages}
    
    Fitting observations of planets and binary stars to orbits has been widely addressed in the literature, dating back to Kepler's seminal work. 
    More recently, a variety of software packages have been released following both frequentist and Bayesian approaches.
    Some of these packages include EXOFAST \citep{eastmanEXOFASTv2PublicGeneralized2019}, rvfit \citep{iglesias-marzoaRvfitCodeDetailed2015}, radvel \citep{fultonRadVelRadialVelocity2018}, and the Python \texttt{exoplanet} package \citep{Foreman-Mackey_exoplanet_Gradient-based_probabilistic_2021}.
    Most relevant to image- and visibility-based modelling are packages for fitting relative astrometry, radial velocity, and proper motion anomaly, like
    \texttt{orbitize!} \citep{bluntOrbitizeComprehensiveOrbitfitting2020}, 
    \texttt{orvara}
    \citep{brandt_orvara_2021_fixed},
    and \texttt{Efit}
    \citep[presented in][]{meyerShortestKnownPeriodStar2012},
    as well as image searching codes like K-Stacker \citep{nowakKStackerKeplerianImage2018} and 
    PACOME (Dallant et al, submitted).
    
    These tools have employed a variety of methods for approximating posteriors, including affine invariant \citep{foreman-mackeyEmceeMCMCHammer2013} and parallel tempered affine invariant Markov chain Monte Carlo \citep{vousdenDynamicTemperatureSelection2016} in  \texttt{orvara} and \texttt{orbitize!},
    Hamiltonian Monte Carlo \citep{hoffmanNouturnSamplerAdaptively2014} in \texttt{exoplanet}, and nested sampling \citep{skillingNestedSampling2004} in \texttt{Efit}.
    With the exception of \texttt{exoplanet}, the majority of these codes have used first-order samplers. That is to say, they rely only on evaluating the posterior density and do not calculate or make use of gradient information.

\section{Results}

\subsection{Demonstration with Relative Astrometry}\label{sec:performance}


    We begin our demonstration of \octo\ by fitting orbits to simulated relative astrometry measurements, which is
    one of the simplest use cases for the package.
    We considered a simulated orbit of a single planet and generated astrometry at 27 epochs (Table \ref{tab:rel-astrom}).
    For the sake of comparison, we performed the same fit using two other popular orbit fitting packages: \texttt{orvara} and \texttt{orbitize!}.

    We followed the best practices laid out in the tutorials provided with each package.
    \texttt{orvara} and \texttt{orbitize!} use slightly different priors by default and cannot be made to match each other exactly without code modifications. 
    \texttt{Orvara} uses an $\sqrt{e}\sin(\omega)$ and $\sqrt{e}\cos(\omega)$ parameterization while \texttt{orbitize!} uses separate uniform priors on both $e$ and $\omega$.
    Therefore, we ran our comparisons twice, first adopting priors similar to \texttt{orvara} and then priors similar to \texttt{orbitize!}
    In all cases, we used a log-uniform prior on semi-major axis between 0.1 and 1000 AU.
    We ran the \texttt{orvara} and \texttt{orbitize!} packages with settings recommended by their authors. These were 4 temperatures with 100 walkers for \texttt{orvara} and 20 temperatures with 1000 walkers for \texttt{orbitize!}.
    We ran \octo\ with 16 independent chains.
    We initialized the walkers used by \texttt{orvara} and \texttt{orbitize!} in a Gaussian ball around the true orbit. For \texttt{orbitize!}, we improved the convergence by randomly initializing half the walkers on the second mode of the $\omega$ marginal posterior.
    For \octo, on the other hand, we initialized it automatically using our procedure of drawing from the priors and selecting the parameters with the highest posterior density.

    We drew 20,000,000 samples from each walker using \texttt{orvara} and 100,000 from each walker using \texttt{orbitize!}.
    For \octo, we adapted the step size and mass matrix for 5,000 iterations and then drew a further 15,000 samples.
    The resulting posteriors are compared in Figure \ref{fig:sampler-comparison}. To remove the burn-in phase, We discarded the first halves of the \texttt{orvara} and \texttt{orbitize!} chains and the first 5,000 samples of the \octo\ chains (the adaptation phase).

    To measure how long each package takes to converge to a steady distribution, we followed a similar procedure to \citet{ferrer-chavezAlgorithmicSpeedupsPosterior2021}. We divided each chain into 50 segments and assumed that in the final segment, the chain is fully converged. We then calculated the $\hat{R}$ statistic between each segment and the final segment. We used the rank normalized and median folded version of the statistic as implemented in \texttt{MCMCDiagnosticTools.jl}\footnote{\url{http://turinglang.org/MCMCDiagnosticTools.jl/}} to evaluate how well the samplers converged in the bulk and in the tails of the distribution. We considered it converged once  $\hat{R}$ became less than $1\pm0.005$ for all variables. 
    We evaluated this on all walkers (\texttt{orvara} and \texttt{orbitize!}) and all chains (\octo) and took the median.

    \begin{figure}
        \includegraphics[width=\colwidth]{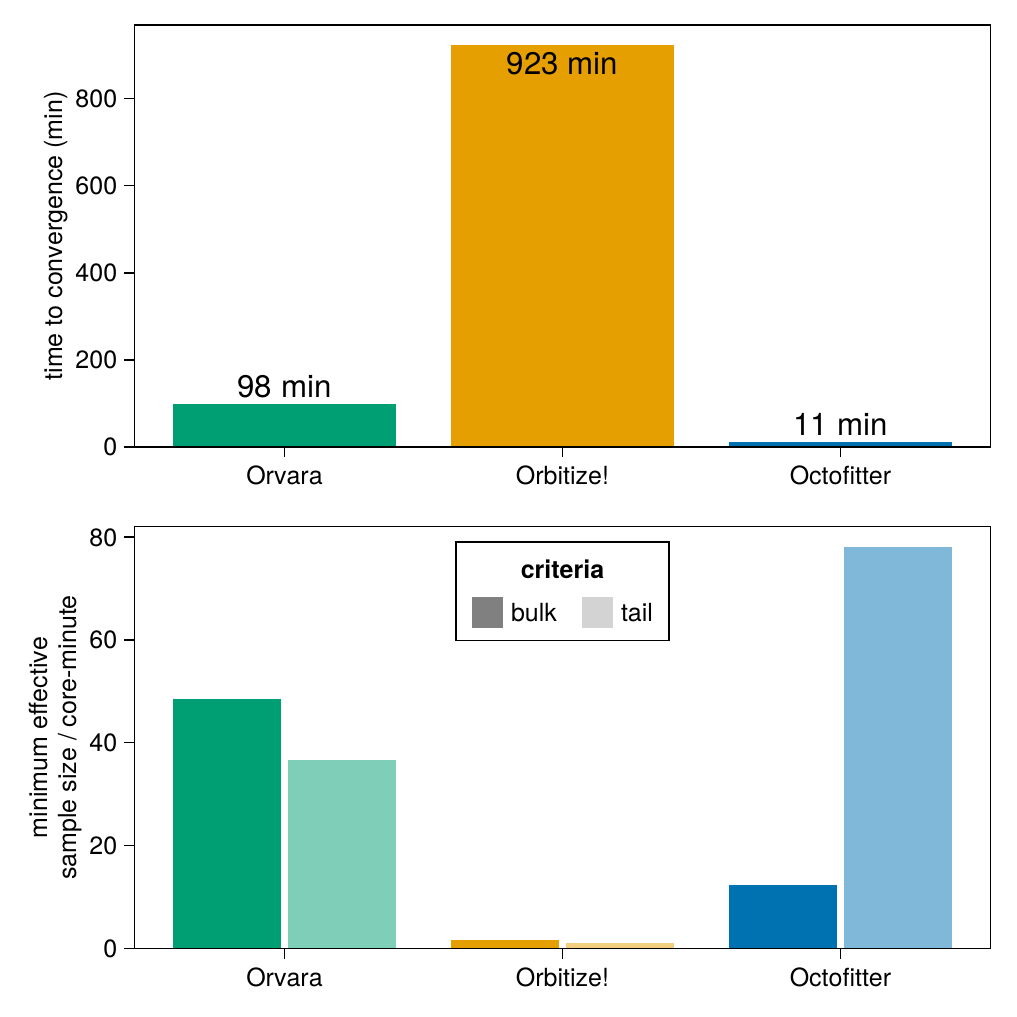}
        \caption{
        Comparison between packages of the average time until chains converged to a stationary distribution and the rate at which independent posterior samples are generated. The effective sample size (ESS) rate was measured separately using the bulk and tail methods of \texttt{MCMCDiagnostics.jl}.
        Note that the $\hat{R}$ and ESS of the slowest variable for a given sampler are used as this is what ultimately limits sampling performance.
        These results are based on the astrometry presented in Table \ref{tab:rel-astrom}, and are expected to depend strongly on hardware, input data, and choice of priors.\label{fig:sampler-efficiency}
        }
    \end{figure}


    The most important result of this code comparison is that in the limit of large numbers of samples,
    all packages produce posterior distributions that largely agree with each other and that include the true orbit. The orbit paths in the plane of the sky are compared in Figure \ref{fig:sampler-comparison} and the parameters are compared in a corner plot in Figure \ref{fig:sampler-comparison-pairplot}.
    This should serve to further improve confidence in the results of all packages.

    \begin{figure*}
        \includegraphics[width=\textwidth]{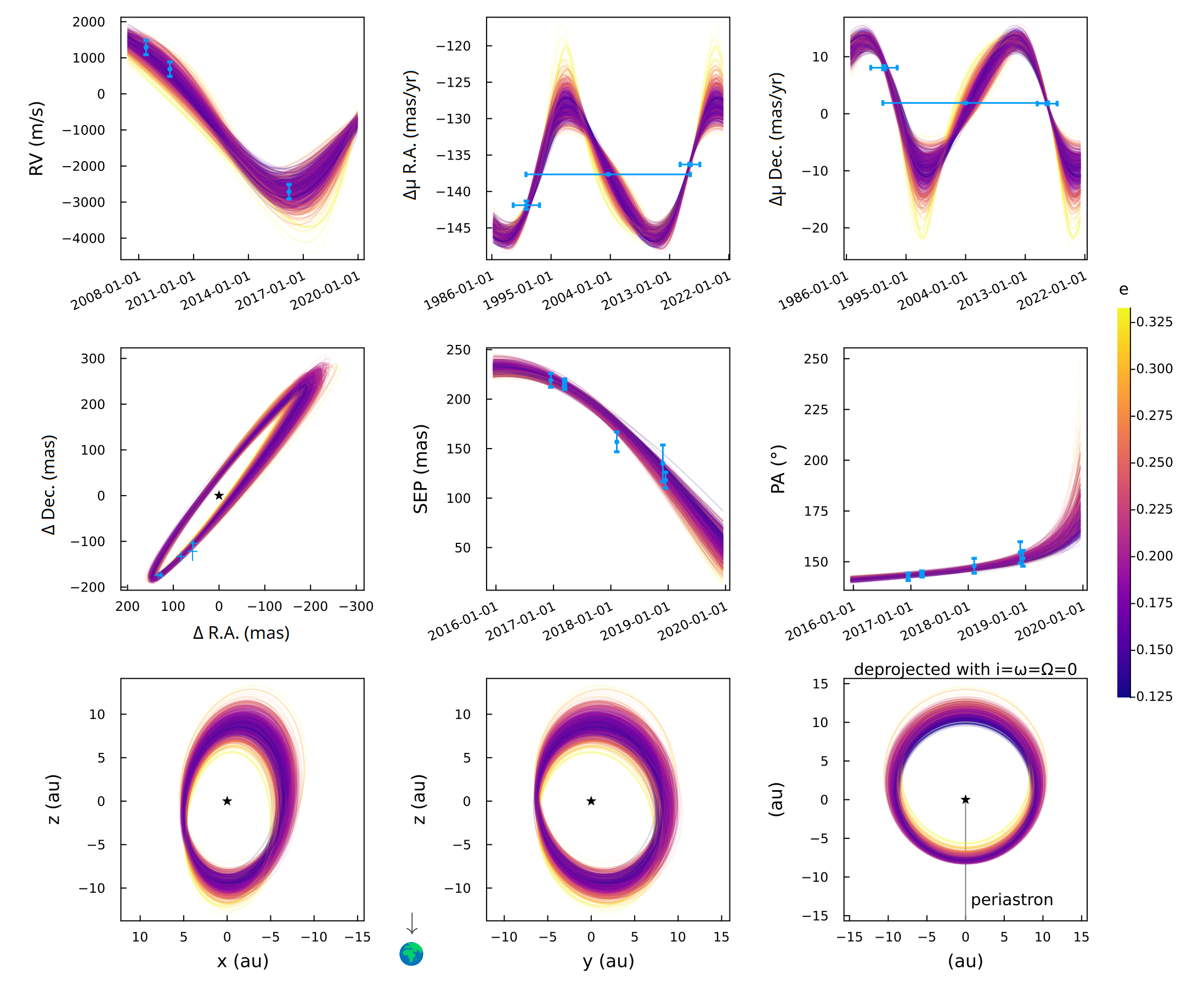}
        \caption{
            Sample plot output from \octo~using data from the HD 91312 system.
            The top row visualizes the orbital posterior compared with velocity measurements of the host.
            The horizontal bars in the proper motion panels show the timespans over which the average velocity was measured.
            The middle row shows the posterior compared with astrometry measurements in the plane of the sky and in separation and position angle over time.
            The bottom row shows the orbital posterior in physical coordinates to compliment the astrometry plot. The rightmost panel shows a deprojected view of the system where orbits have been rotated face on and to place periastron at the bottom. The conventions used by \octo~are described in Appendix \ref{sec:derivation}.
            \label{fig:hd91312}
        }
    \end{figure*}

    The sole exception is that the orbit posterior from \octo\ contains more samples from a high-eccentricity, high-semi-major axis branch of the posterior than the two other packages (Figure \ref{fig:sampler-comparison-pairplot}). Nonetheless, running SBC on this model with uncertainties and observation epochs from this dataset (Figure \ref{fig:sbc-grid}) reveals that \octo\ is acceptably calibrated and is not overestimating the spread of the posterior. It is not computationally feasible to perform the same SBC experiment with the other packages, but this likely indicates that the additional samples from the long tail of the posterior are representative of the true posterior and would eventually appear in the outputs of the other packages.
    
    A second result is that for this problem,
    \octo\ converges to a steady distribution almost immediately using a single computer core, while other packages take many hours to converge (Figure \ref{fig:sampler-efficiency}).
    Since assessing the convergence of MCMC chains can be difficult, we suggest that with \octo, unsophisticated users are therefore less likely to use insufficiently converged results in their analyses.
    Of course, in addition to needing the chain to be converged, one must also generate a sufficient number of samples for the problem at hand.
    Interestingly, despite the much faster convergence, \octo\ is no faster than the other packages at producing independent samples. 
    In most workflows, the bottleneck is ensuring convergence rather than a need to produce tens of thousands of samples, so \octo\ should still offer a decisive improvement in computation time.
    
    The exact results are hardware specific,
    sensitive to the data,
    and depend on the choice of priors and parameterization.
    It is not feasible to fully account for all small differences between software packages,
    and it is likely that different approaches will perform better or worse depending on the problem at hand. 
    An additional caveat is that the \texttt{ptemcee} sampler used by \texttt{orvara} and \texttt{orbitize!} scales across many cores which reduces the total sampling time.
    \octo\ supports running multiple chains in parallel but is not configured to use multiple cores to accelerate a single chain.
    Finally, given the convergence guarantees of Markov chain Monte Carlo, all packages would  approach the same distribution given infinite time.
    This comparison, therefore, is meant only to illustrate the typical efficiency one might expect with \octo~on similar problems and reasonable run times.

    \subsection{Demonstration with Relative Astrometry, RV, and PMA}
    
    We now demonstrate \octo~on a real system with radial velocity, proper motion anomaly, and relative astrometry data.

    The HD 91312 system consists of a $1.6 \mathrm{M_{\odot}}$ star orbited by a binary companion with a mass of approximately $0.3 \mathrm{M_{\odot}}$ and separation of roughly 10 AU \citep{chilcoteSCExAOCHARISDirect2021}.
    The companion was discovered by a direct imaging search targeting accelerating stars \citep{currieNewTypeExoplanet2021} from the HGCA.
    We now reproduce the orbital analysis of the discovery paper in order to demonstrate \octo's~modelling capabilities when applied to a combination of relative astrometry, radial velocity, and proper motion anomaly data.
    
    For this analysis, we used the astrometry data from \citet{chilcoteSCExAOCHARISDirect2021},
    proper motion anomaly data from the HGCA \citep{brandtHipparcosGaiaCatalogAccelerations2021},
    and limited radial velocity data originally from \citet{borgnietExtrasolarPlanetsBrown2019}.
    As the original radial velocity data was not forthcoming, we measured it from the PDF plots submitted to the arXiv alongside the 2021 manuscript.
    We used similar priors as those described in the discovery paper, namely log-normal priors on primary and companion mass and a uniform prior on the square root of eccentricity.
    These were chosen to match the analysis of the discovery paper for the purpose of comparing codes and demonstrating \octo\ rather than any physical motivation on our part.

    The results of this orbit modelling are presented in Figure \ref{fig:hd91312}.
    \octo~recovers the orbit of the companion with similar results to those presented by \citet{chilcoteSCExAOCHARISDirect2021}.
    The radial velocity, proper motion anomaly, and relative astrometry are all consistent with the secondary companion having a mass of approximately of 300 $\mathrm{M_{jup}}$,
    given the choice of priors described in the original discovery paper.

    \subsection{Demonstration with Images}

    
    \begin{figure*}
        \centering
        \includegraphics[width=0.8\textwidth]{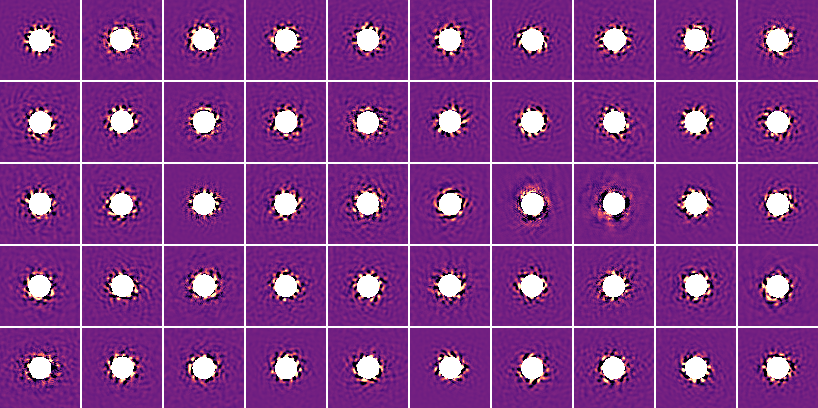}
        \caption{GPI sequences used for simulations in this section. Each image was normalized to have the same average contrast at 200 mas separation from the star (just outside the edge of the mask). The images displayed above contain a simulated planet orbiting CCW at an average of just SNR 1 per epoch, spaced one month apart. Given this data, the model recovers the simulated planet at SNR 7.}
        \label{fig:gpi-image-mosiac}
    \end{figure*}

        

    \begin{figure*}
        \centering
        \includegraphics[width=0.45\textwidth]{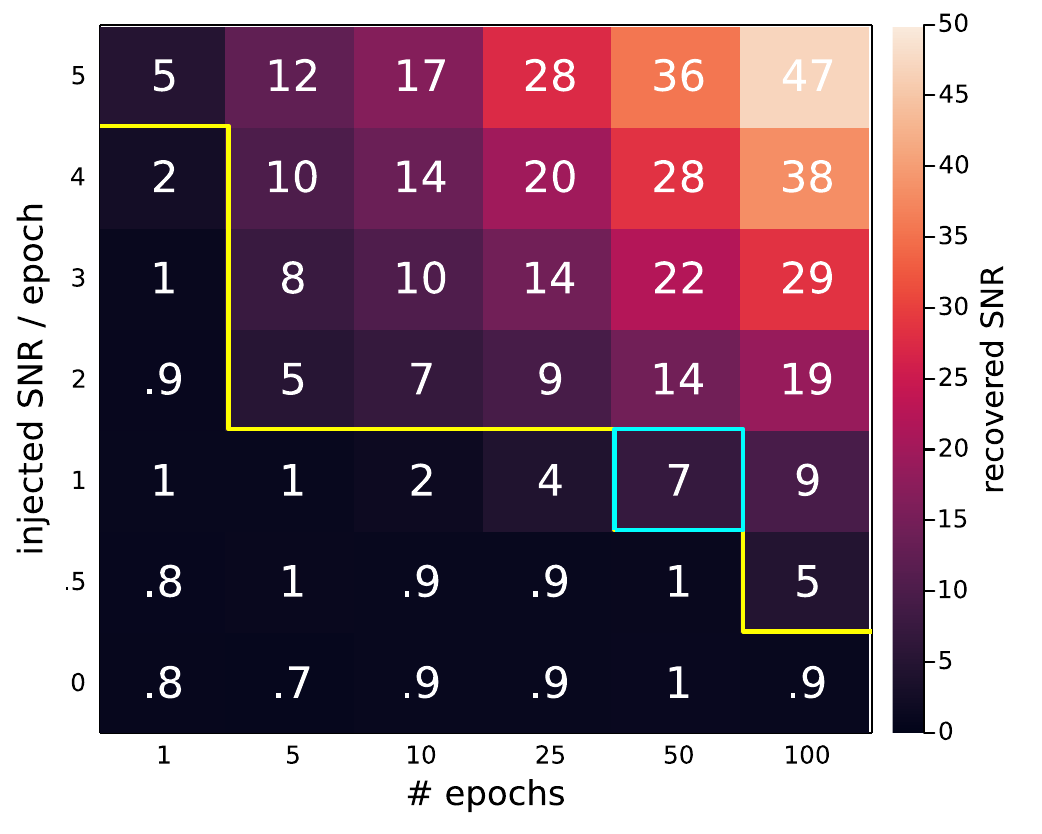}
        \includegraphics[width=0.45\textwidth]{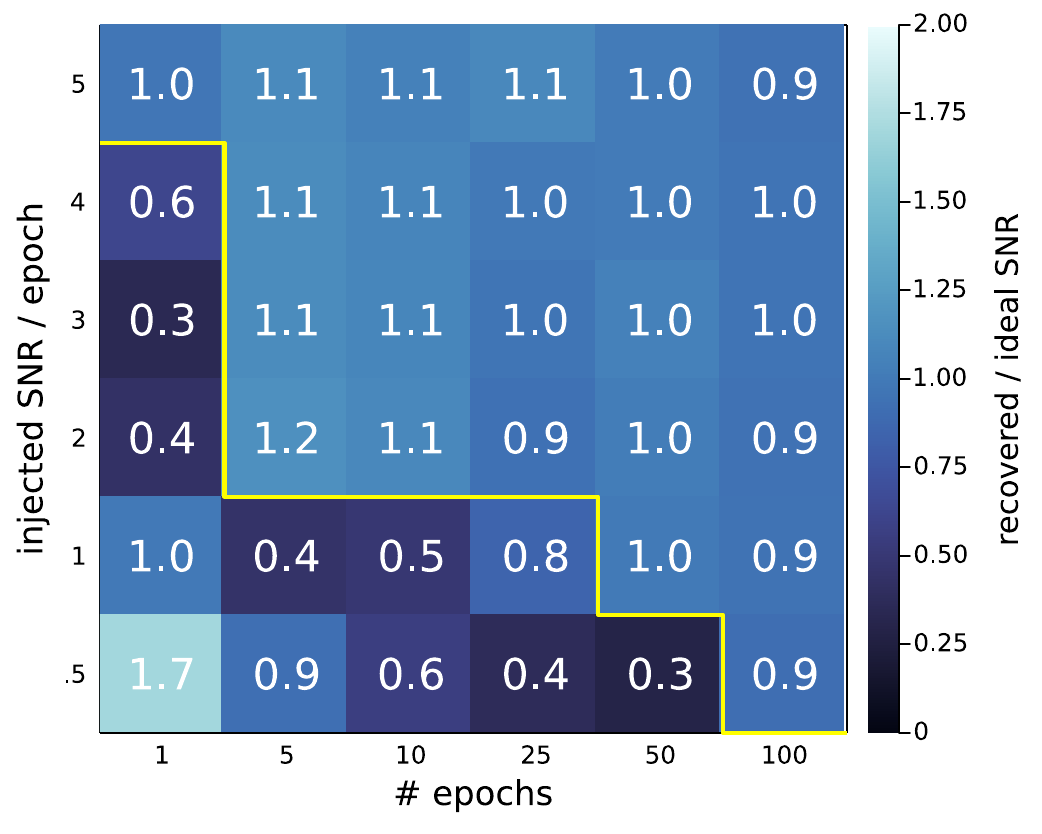}
        \caption{\octo's ability to recover planets from simulations of circular, face-on orbits as a function of signal to ratio per epoch and number of epochs. 
        \textbf{Left:} Recovered SNR. Cells above the yellow line would be detected with a $5\sigma$ threshold. The cell outlined in cyan corresponds to the data shown in Figure \ref{fig:detection-example}.
        \textbf{Right:} The same SNRs relative to an ideal $\sqrt{N}$ improvement with the number of epochs.
        We find that the SNR grows as expected unless the final, combined SNR is below $\approx5$.
        The recovered SNR falls off quickly below this value and levels off at approximately 1. This explains why the recovered SNR of 1 exceeds the injected value of 0.5 in the bottom left corner.
        \label{fig:circular-face-on-recovered-snr}}
    \end{figure*}
        
    We now present a series of simulations showing how this framework allows us to detect planets using multi-epoch direct images.
    We selected 50 sequences from the Gemini Planet Imager Exoplanet Survey \citep{nielsenGeminiPlanetImager2020}, processed using a forward model matched filter \citep[FMMF,][]{ruffioImprovingAssessingPlanet2017}, that have a stellar I-band magnitude less than or equal to 6 and have no previously detected point sources.
    To be clear, we do not search these sequences for real companions but merely use them as realistic noise maps for our simulations.
    We normalized the contrasts of each sequence to the median at 200 mas separation. This retains the true noise distribution but removes the effects of sequence-to-sequence variation on our results.
    We consider a hypothetical system at $20 \; \mathrm{pc}$ with a 
    $1 \mathrm{M_{\odot}}$
    star that is observed once per month for between 1 and 100 months (a little over eight years).
    Using these sequences and parameters, we generated simulated observations of a planet by injecting a synthetic PSF into the correct positions in each epoch.
    This input data is presented in Figure \ref{fig:gpi-image-mosiac}.
    
    For all models, we adopted a $1\pm 0.1 \mathrm{M_{\odot}}$ prior on $M$, a uniform $0-30 \mathrm{M_{jup}}$ prior on $m$, a Gaussian prior on parallax, a uniform prior on $a$, and a uniform circular prior on $\tau$.
    

    We test this model's ability to detect planets in sequences of direct images for the most straightforward case: circular, face-on orbits.
    We injected the planet into 1, 5, 10, 25, 50, and 100 images with an average SNR ranging between 0 and 5 in each image.
    Finally, we applied our model to each of these simulated datasets to arrive at a grid of recovered SNR values as a function of number of epochs and SNR per epoch (Figure \ref{fig:circular-face-on-recovered-snr}).
    We find that we are able to recover planet detections with arbitrarily low SNR per epoch despite orbital motion, provided we have a sufficient number of observations.

    Figure \ref{fig:circular-face-on-recovered-snr} also shows how the recovered SNR compares with the SNR we would expect for combining images without orbital motion in the presence of uncorrelated Gaussian noise.
    We find that the model detects the injected planets with near perfect $\sqrt{N}$ scaling when the final, combined SNR is greater than $\sim$5.
    For instance, a planet injected into 100 epochs spaced one month apart, at an SNR per epoch of just 0.5 (below the noise), is still robustly detected at a final SNR of 5.
    We do note that since each image was taken of a different star, this ideal scaling is a best-case scenario. It is possible that repeated observations of the same target could lead to correlated residuals that reduce the final SNR, though the orbital motion of the planet should mitigate this in much the same way as angular differential imaging \citep{malesOrbitalDifferentialImaging2015,maroisAngularDifferentialImaging2006}.
    
    The left-hand columns of Figure \ref{fig:circular-face-on-recovered-snr} are of particular note. They show that the model fails to detect a point source injected into a single image at an SNR of 4.
    The failure to detect planets with expected significance below SNR $\approx 5$ can be understood by contrasting our detection criteria with the standard used in the field.
    In these models, we consider the overall SNR marginalized over all locations in the image (or orbits through a sequence), whereas typically, one looks at the maximum SNR at any given location.
    For example, in any SNR 5 detection of a planet, numerous other SNR 2 and 3 peaks exist. When looking at the SNR marginalized over all locations, these other peaks serve to reduce the final SNR.
    This makes the SNR calculated from the marginal photometry posterior a more stringent planet detection test.

\begin{figure*}[p]
    \includegraphics[width=\textwidth]{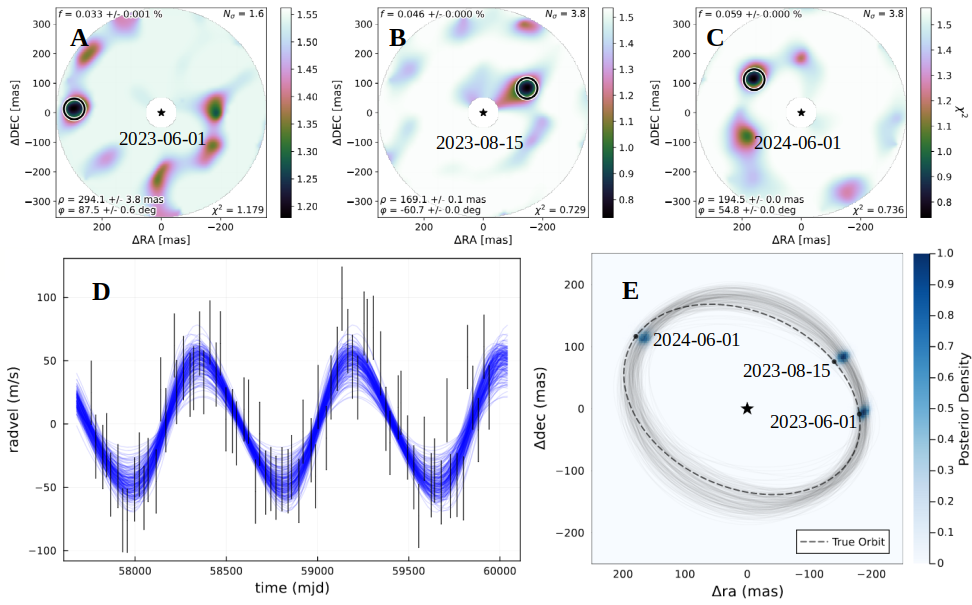}
    \caption{Joint modelling of AMI and RV data. \textbf{A-C}: Recovered $\chi^2$ maps at each independent AMI epoch created using \texttt{Fouriever} \citep[presented in][]{kammererInfraredImagerSlitless2023}. \textbf{D and E:} joint radial velocity and AMI posteriors visualized as an RV time series (D) and orbit posterior in the plane of the sky (E).
    \label{fig:visibility-plot}}
\end{figure*}

\subsection{Demonstration with Aperture Masking Interferometry and Radial Velocity}

We now demonstrate \octo\ using a combination of simulated radial velocity data and aperture masking interferometry (AMI) visibilities.
We considered a plausible scenario where a planet has been detected using radial velocity, had its orbit characterized, and is then followed up with a series of three observations with JWST NIRISS AMI \citep{doyonJWSTFineGuidance2012, sivaramakrishnanInfraredImagerSlitless2023}.
For this experiment, we connected the mass of the planet to its $Ms$ band photometry using Sonora Bobcat models \citep{marleySonoraBobcatCloudfree2021} for a fixed system age of 10 Myr.
The simulated system had a true orbit with a semi-major axis of 2 AU, an eccentricity of 0.1, an inclination of 45$\degr$, a parallax distance of 100 mas, and a mass of 5 $\mathrm{M_{jup}}$. The star had a stellar absolute $Ms$-band magnitude of 2.6.

We simulated the radial velocity by adding Gaussian noise with a standard deviation of 25 m/s to 74 epochs of RV data generated from the above system. The data points were spaced by $\sim$30 days between 2017 and 2023. 

The AMI data was generated using Eq. \ref{eq:vis-bin} for a single companion following the given orbit.
For this example, we used only the closure phase and did not include the squared visibilities.
We again added Gaussian noise to the calculated closure phases, with a closing triangle dependent standard deviation taken from the closure phase uncertainties calculated with \texttt{AMICAL} \citep{2020SPIE11446E..11S} from NIRISS AMI F480M data of (the presumed single stars) HD 101531 and HD 123991 calibrated against each other. 

We applied \octo\ with a joint model of the RV data and three AMI epochs. We also compared this to $\chi^2$ maps from the \texttt{Fouriever} framework \citep[presented in][]{kammererInfraredImagerSlitless2023} of each individual epoch.
The results of this experiment are shown in Figure \ref{fig:visibility-plot}.

Using the AMI data, the \octo\ model is able to constrain the inclination and, therefore, the mass of the planet. Due to the faint signal and presence of noise, the \texttt{Fouriever} results show spurious signals at each epoch. In comparison, \octo\ is able to connect the three epochs by a single higher-significance mode in orbit-parameter space.

This example illustrates how joint modelling across epochs can be used to increase the significance of AMI and other similar interferometric observations.

\section{Conclusion}

We have presented a new code, \octo, for modelling exoplanet relative astrometry, radial velocity, and proper motion anomaly, as well as performing non-traditional tasks like detecting moving point sources across images and interferometric observables.
\begin{itemize}
    \item We demonstrated the simulation-based calibration procedure on a hypothetical orbit fitting task and found that for orbits parameterized with Thiele-Innes constants, \octo\ is acceptably calibrated.
    \item We compared the results of \octo\ to the popular packages \texttt{orvara} and \texttt{orbitize!} and found that all three arrive at similar posterior distributions.
    \item We showed that \octo\ can converge 1 to 2 orders of magnitude faster than other popular packages, making it computationally feasible to perform a variety of statistical checks like simulation-based calibration for individual datasets.
    \item We demonstrated a combined fit to relative astrometry, radial velocity, and proper motion anomaly of the HD 91312 system and found a companion mass that agreed with previous results.
    \item We demonstrated the ability to detect arbitrarily faint companions despite orbital motion using simulations and data from the Gemini Planet Imager Exoplanet Survey.
    \item We demonstrated a combined model of simulated radial velocity and multi-epoch JWST/NIRISS aperture masking interferometry.
\end{itemize}
\octo\ is a powerful new tool that will enable the community to broadly apply joint multi-epoch, multi-instrument, and multi-band direct imaging, interferometry, Doppler radial velocity, and astrometric motion. This will allow for the detection and characterization of fainter and lower-mass planets more similar to those in our solar system.

\vspace{5mm}
\facilities{Gemini (GPI), Gaia, Hipparcos}


\software{Julia \citep{bezansonJuliaFastDynamic2012}, Makie \citep{danischMakieJlFlexible2021}, GR \citep{GRFramework}}




\appendix

\section{Derivation of Cartesian and Celestial Position, Velocity, and Acceleration} \label{sec:derivation}


To properly fit to relative astrometry, proper motion anomaly, and radial velocity data, expressions for the position, velocity, and acceleration of a secondary companion (such as a planet) about a primary mass (such as a star) in Cartesian and celestial coordinates are required.
It is well-known that the solution to the Keplerian two-body problem \citep[for example][]{goldsteinClassicalMechanics2008} is
\begin{equation}
    r(t) = \frac{a(1 - e^2)}{1 + e\cos(\nu(t))},
\end{equation}
where $r$ is the radial separation of the two bodies, $a$ is the orbital semi-major axis, $e$ is the orbital eccentricity, and $\nu(t)$ is the true anomaly.
The true anomaly describes the angular position of the orbit with respect to periastron, and is the solution to the equation
\begin{equation}
    \tan\left( \frac{\nu(t)}{2} \right) = \sqrt{\frac{1 + e}{1 - e}} \tan\left( \frac{E(t)}{2} \right). \label{eq:trueanomaly}
\end{equation}
The eccentric anomaly $E(t)$ is in turn the solution to
\begin{equation}
    M(t) = E(t) - e\sin(E(t)), \label{eq:eccentricanomaly}
\end{equation}
with $M(t)$ being the mean anomaly
\begin{equation}
    M(t) = \frac{2\pi(t - t_{\mathrm{peri}})}{T}. \label{eq:meananomaly}
\end{equation}
Here, $t_{\mathrm{peri}}$ is the time of periastron passage and $T$ is the orbital period.
Since Eq. \ref{eq:eccentricanomaly} is a transcendental equation, $E(t)$ must be determined numerically.
Once calculated, though, the derivatives of the eccentric anomaly with respect to the mean anomaly and eccentricity have analytic expressions:
\begin{equation}
    \frac{\partial E}{\partial M} = \frac{1}{1 - e \cos{E(t)}} \label{eq:kepler-partial-ma}
\end{equation}
and
\begin{equation}
    \frac{\partial E}{\partial e} = \frac{\sin{E(t)}}{1 - e \cos{E(t)}}. \label{eq:kepler-partial-e}
\end{equation}
These expressions can be determined using implicit differentiation.

\subsection{Cartesian Coordinates \label{sec:cartesian-coordinates}}
Using the well-known relation between polar coordinates $(r, \nu)$ and Cartesian coordinates $(x, y, z)$, we see that in the frame of the secondary, its position with respect to the primary is
\begin{equation}
    \mathbf{r}(t) = \begin{bmatrix} r(t)\cos(\nu(t)) \\ r(t)\sin(\nu(t)) \\ 0 \end{bmatrix} = \frac{a(1 - e^2)}{1 + e\cos(\nu(t))} \begin{bmatrix} \cos(\nu(t)) \\ \sin(\nu(t)) \\ 0 \end{bmatrix}.
\end{equation}
However, the frame of the secondary and the frame of an external observer are unlikely to be coincident.
Instead, an external observer is likely to be rotated with respect to the frame of the secondary.
This rotation is quantified by $\omega$ (the argument of periastron of the secondary), $i$ (the inclination of the orbit), and $\Omega$ (the longitude of the orbit's ascending node).
The rotations associated with these angles are given by the rotation operators
\begin{equation}
    R_z(\omega) = \begin{bmatrix} \cos(\omega) & -\sin(\omega) & 0 \\ \sin(\omega) & \cos(\omega) & 0 \\ 0 & 0 & 1 \end{bmatrix}, \quad R_x(i) = \begin{bmatrix} 1 & 0 & 0 \\ 0 & \cos(i) & -\sin(i) \\ 0 & \sin(i) & \cos(i) \end{bmatrix}, \quad R_z(\Omega) = \begin{bmatrix} \cos(\Omega) & -\sin(\Omega) & 0 \\ \sin(\Omega) & \cos(\Omega) & 0 \\ 0 & 0 & 1 \end{bmatrix}.
\end{equation}
Thus, the position of the secondary with respect to its primary as seen by an external observer is
\begin{equation}
    \mathbf{r}_{\text{obs}}(t) = R_z(\Omega)R_x(i)R_z(\omega)\mathbf{r}(t) = r(t)\begin{bmatrix} \cos(\nu(t) + \omega)\cos(\Omega) - \sin(\nu(t) + \omega)\cos(i)\sin(\Omega) \\ \cos(\nu(t) + \omega)\sin(\Omega) + \sin(\nu(t) + \omega)\cos(i)\cos(\Omega) \\ \sin(\nu(t) + \omega)\sin(i) \end{bmatrix}.
\end{equation}
Explicitly, we have
\begin{equation}
    \mathbf{r}_{\text{obs}}(t) = \frac{a(1 - e^2)}{1 + e\cos(\nu(t))}\begin{bmatrix} \cos(\nu(t) + \omega)\cos(\Omega) - \sin(\nu(t) + \omega)\cos(i)\sin(\Omega) \\ \cos(\nu(t) + \omega)\sin(\Omega) + \sin(\nu(t) + \omega)\cos(i)\cos(\Omega) \\ \sin(\nu(t) + \omega)\sin(i) \end{bmatrix} \label{eq:robs}
\end{equation}

Inspecting this equation, we see that $\mathbf{r}_{\text{obs}}(t)$ is dependent on $a$, $e$, $i$, $\omega$, and $\Omega$.
In addition, there is dependence on $t_{\mathrm{peri}}$ and $T$ because of Eqs. \ref{eq:trueanomaly} and \ref{eq:eccentricanomaly}.
However, $a$ and $T$ are related by Kepler's third law, meaning dependence on $T$ can be reformulated as dependence on $a$.
Thus, a Keplerian orbit is uniquely described by the parameters $(a, e, i, \omega, \Omega, t_{\mathrm{peri}})$; these are referred to as the Campbell elements.
Note that in Octofitter, we replace $t_{\mathrm{peri}}$ with either $\tau$ or $\theta$. $\tau$ is given by
\begin{equation}
    \tau = \left( \frac{t_{\mathrm{peri}} - t_{\text{ref}}}{T} \right) \mod 1,
\end{equation}
where we choose $t_{\mathrm{ref}} = 58849.0$ MJD (January 1, 2020) \citep{bluntOrbitizeComprehensiveOrbitfitting2020}, and $\theta$ is discussed in Appendix \ref{sec:theta-tau-derivation}.
We now differentiate $\mathbf{r}_{\text{obs}}(t)$ to obtain $\mathbf{v}_{\text{obs}}(t)$.
Kepler's second law implies that
\begin{equation}
    \frac{r(t)^2\dot{\nu}(t)}{2} = \frac{\pi a^2 \sqrt{1 - e^2}}{T}.
\end{equation}
From this, we obtain
\begin{equation}
    \dot{r}(t) = \frac{2\pi a}{T} \frac{e\sin(\nu(t))}{\sqrt{1 - e^2}}, \quad r(t)\dot{\nu}(t) = \frac{2\pi a}{T} \frac{1 + e\cos(\nu(t))}{\sqrt{1 - e^2}}.
\end{equation}
Using these identities to differentiate $\mathbf{r}_{\text{obs}}(t)$ produces
\begin{equation}
    \mathbf{v}_{\text{obs}}(t) = \frac{2\pi a}{T} \frac{1}{\sqrt{1 - e^2}} \begin{bmatrix} -\cos(i)\sin(\Omega)[\cos(\nu(t) + \omega) + e\cos(\omega)) + \cos(\Omega)(\sin(\nu(t) + \omega) + e\sin(\omega)] \\ \cos(i)\cos(\Omega)[\cos(\nu(t) + \omega) + e\cos(\omega)) - \sin(\Omega)(\sin(\nu(t) + \omega) + e\sin(\omega)] \\ \sin(i)[\cos(\nu(t) + \omega) + e\cos(\omega)] \end{bmatrix}. \label{eq:vobs}
\end{equation}
We determine $\mathbf{a}_{\text{obs}}(t)$ in an identical manner.
Observing that
\begin{equation}
    \dot{\nu}(t) = \frac{2\pi}{T} \frac{(1 + e\cos(\nu(t)))^2}{(1 - e^2)^{3/2}},
\end{equation}
differentiating $\mathbf{v}_{\text{obs}}(t)$ yields
\begin{equation}
    \mathbf{a}_{\text{obs}}(t) = \frac{4\pi^2 a}{T^2} \frac{1}{(1 - e^2)^2} \begin{bmatrix} (1 + e\cos(\nu(t)))^2[\cos(i)\sin(\Omega)\sin(\nu(t) + \omega) - \cos(\Omega)\cos(\nu(t) + \omega)] \\ -(1 + e\cos(\nu(t)))^2[\cos(i)\cos(\Omega)\sin(\nu(t) + \omega) + \sin(\Omega)\cos(\nu(t) + \omega)] \\ -(1 + e\cos(\nu(t)))^2\sin(i)\sin(\nu(t) + \omega) \end{bmatrix}. \label{eq:aobs}
\end{equation}
Having expressions for $\mathbf{r}_{\text{obs}}(t)$, $\mathbf{v}_{\text{obs}}(t)$, and $\mathbf{a}_{\text{obs}}(t)$ is quite useful, since they allow us to determine the corresponding quantities in celestial coordinates with ease.
Furthermore, the radial velocity of the secondary with respect to its primary can be read off directly as the $z$-component of Eq. \ref{eq:vobs}, giving us
\begin{equation}
    v_{z,\text{obs}} = \frac{2\pi a}{T}\frac{\sin(i)[\cos(\nu(t) + \omega) + e\cos(\omega)]}{\sqrt{1 - e^2}}.
\end{equation}
To get the primary's radial velocity with respect to the secondary, we simply employ conservation of momentum and multiply the previous expression by $-M_{\text{s}}/(M_{\text{s}} + M_{\text{p}})$, with $M_{\text{p}}$ and $M_{\text{s}}$ being the masses of the primary and secondary, respectively.
To determine the radial velocities measured by an external observer, we add the barycentre radial velocity $\gamma$. Additional linear trends, quadratic trends, or other trends can be accounted for in \octo\ by making the instrument zero-points a function of other parameters in the fit.
Overall, we get that the observed radial velocities of the primary and secondary are
\begin{align}
    v_{\text{r},\text{s}} &= \frac{2\pi a}{T}\frac{\sin(i)[\cos(\nu(t) + \omega) + e\cos(\omega)]}{\sqrt{1 - e^2}} + \gamma + \xi, \\
    v_{\text{r},\text{p}} &= -\frac{M_{\text{s}}}{M_{\text{s}} + M_{\text{p}}}\frac{2\pi a}{T}\frac{\sin(i)[\cos(\nu(t) + \omega) + e\cos(\omega)]}{\sqrt{1 - e^2}} + \gamma + \xi,
\end{align}
where $\xi$ is the combined effect of all relevant trends.
Using these expressions, orbital and physical parameters can be fit to radial velocity data.

\subsection{Celestial Coordinates}
Now that we know $\mathbf{r}_{\text{obs}}(t)$, $\mathbf{v}_{\text{obs}}(t)$, and $\mathbf{a}_{\text{obs}}(t)$, we can calculate the corresponding expressions in celestial coordinates.
Before doing so, a brief discussion of coordinate conventions is required.
For this paper, we measure with reference to the celestial north pole.
This means that, in reference to the previous section, the positive $x$-axis points in the direction of positive declination (upwards on the sky), while the positive $y$-axis points in the direction of positive right ascension (leftwards on the sky).
This choice of coordinates is important to note, since it differs from the right/up orientation of the $x$/$y$ axes commonly seen elsewhere. Furthermore, the positive $z$-axis points away from the observer.

Keeping this coordinate convention in mind, we can determine position, velocity, and acceleration in celestial coordinates.
Normally, an observer a distance $d$ away from an object of size $r$ measures the object to have an angular size of
\begin{equation}
    \Delta \theta = \arctan \left( \frac{r}{d} \right).
\end{equation}
However, for scenarios relevant to this work, $r$ is on the order of astronomical units and $d$ is on the order of parsecs, meaning the small angle approximation $\Delta \theta \approx r/d$ can be used with negligible error.
Applying this approximation, we get that the right ascension and declination offsets of the secondary from its primary are

\begin{equation}
    \Delta\mathrm{RA} = \frac{y_{\text{obs}}}{d}, \quad \Delta\mathrm{DEC} = \frac{x_{\text{obs}}}{d},
\end{equation}
where $d$ is the distance to the system, and $x_{\text{obs}}$ and $y_{\text{obs}}$ are the $x$- and $y$-components of Eq. \ref{eq:robs}.
From this, it immediately follows that
\begin{equation}
    \Delta\dot{\mathrm{RA}} = \frac{v_{y,\text{obs}}}{d}, \quad \Delta\dot{\mathrm{DEC}} = \frac{v_{x,\text{obs}}}{d},
\end{equation}
where $v_{x,\text{obs}}$ and $v_{y,\text{obs}}$ are the $x$- and $y$-components of Eq. \ref{eq:vobs}, and that
\begin{equation}
    \Delta\ddot{\mathrm{RA}} = \frac{a_{y,\text{obs}}}{d}, \quad \Delta\ddot{\mathrm{DEC}} = \frac{a_{x,\text{obs}}}{d},
\end{equation}
where $a_{x,\text{obs}}$ and $a_{y,\text{obs}}$ are the $x$- and $y$-components of Eq. \ref{eq:aobs}.
The equations for $\Delta\mathrm{RA}$ and $\Delta\mathrm{DEC}$ can be used to fit orbital and physical parameters to relative astrometry data, while the equations for $\Delta\dot{\mathrm{RA}}$ and $\Delta\dot{\mathrm{DEC}}$ can be used to fit orbital and physical parameters to proper motion anomaly data.
$\Delta\ddot{\mathrm{RA}}$ and $\Delta\ddot{\mathrm{DEC}}$ are used for calculating model derivatives for higher-order samplers, and may be applied to angular acceleration data in the future.

    \begin{figure*}
        \includegraphics[width=\textwidth]{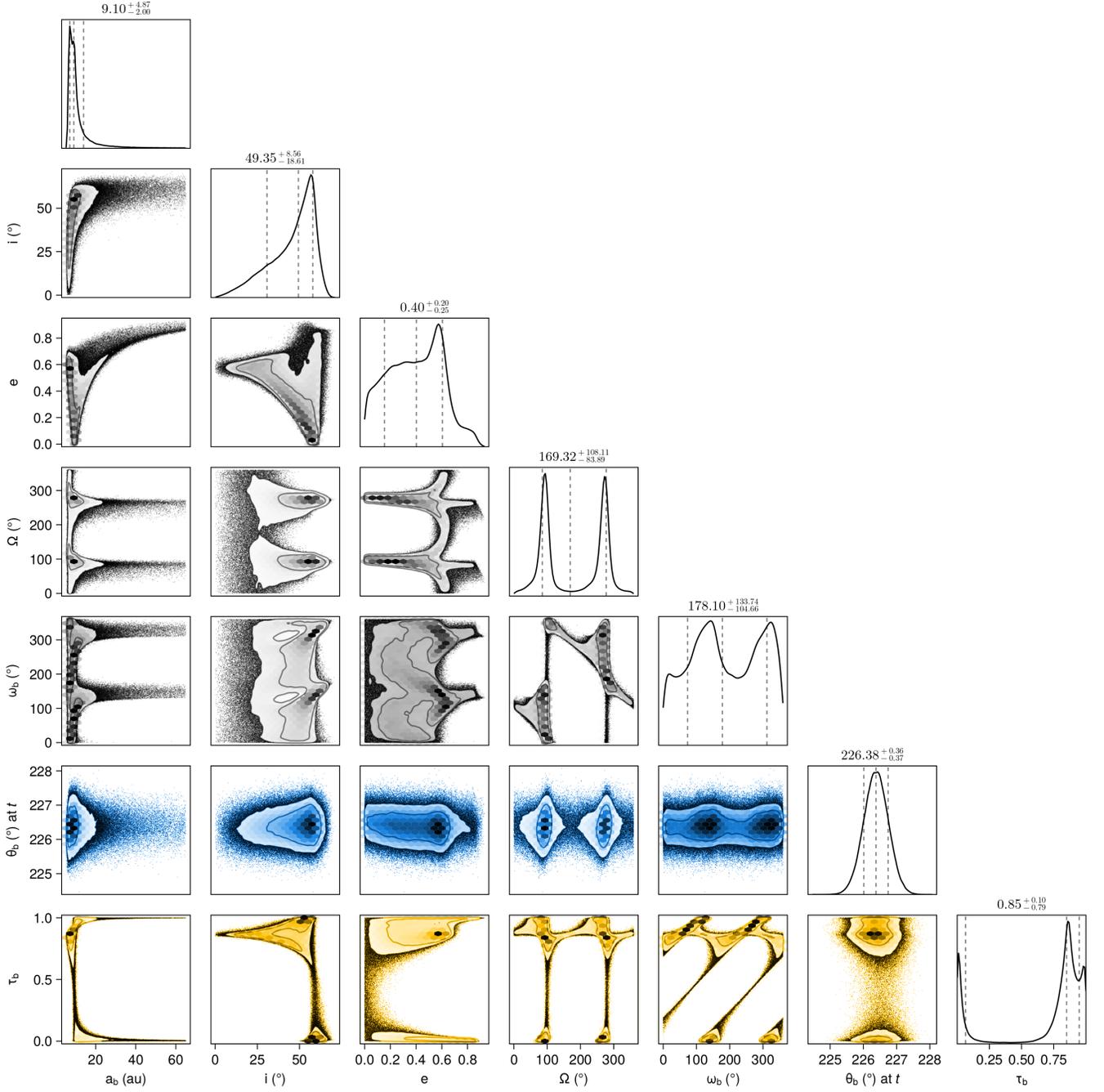}
        \caption{Comparison of $\theta$ (blue, second last row) and  $\tau$ (yellow, last row) paramterizations of the same orbital posterior.
        The $\theta$ marginal posterior is nearly Gaussian, and has simple relationships with all other orbital parameters. By contrast, the $\tau$ marginal posterior is less informative and has complex relationships to $\omega$ and $\Omega$.
        Complex structures in a posterior decrease sampling efficiency and could in some cases lead to biased results.
            \label{fig:pairplot-octo-tau-theta}}
    \end{figure*}

\section{Derivation of the Parameter $\tau$ from Position Angle $\theta$} \label{sec:theta-tau-derivation}

The parameter $\tau$ used in \texttt{orbitize!} and \octo\ is the fraction of the orbit completed by a planet after periastron, at some reference epoch $t_\mathrm{ref}$.
This is a useful parameterization since it lies between 0 and 1 for all orbits; however, if the reference epoch is not similar to the epochs of the observations, it can exhibit pathological behaviour as the other orbital parameters change. 
One solution is to adjust the reference epoch for each dataset, but a further improvement for many cases is to instead adopt the observed position angle $\theta$ at epoch $t$ as an independent variable.
$\theta$ is an improvement over $\tau$  because it is insensitive to changes in the other parameters and has a straightforward interpretation. 
A derivation of $\tau$ from $\theta$ is given here.

We begin with the parallax distance to the system $\bar{\omega}$, the host mass $M$, the eccentricity $e$, and the Thiele-Innes constants $A, B, F,$ and $G$ given in Eqs. \ref{eq:thiele-innes-A} -- \ref{eq:thiele-innes-G}.
From these, we write the transformation matrix
\begin{equation}
   T =  \begin{bmatrix} A & B \\ F & G \end{bmatrix},
\end{equation}
from which we find the unit vectors 
\begin{equation}
   \begin{bmatrix} \hat{x} \\ \hat{y} \end{bmatrix} = T^{-1} \begin{bmatrix} \sin(\theta) \\ \cos(\theta) \end{bmatrix}.
\end{equation}
We can then calculate true anomaly $\nu$ as
\begin{equation}
   \nu = \arctan(\hat{y},\hat{x}),
\end{equation}
and mean anomaly as
\begin{equation}
    \mathrm{MA} = \arctan(-\sin(\nu)\sqrt{1 - e^2}, -\cos(\nu) - e) + \pi - \frac{e\sqrt{1 - e^2}\sin(\nu)}{1 + e\cos(\nu)}.
\end{equation}

From this, the parameter $\tau$ can be calculated using the period as follows.
First, calculate the semi-major axis $a$ using Eq. \ref{eq:a-from-uv}. Next, combining Kepler's third law and the value of $a$ gives us the period $P = 2\pi\sqrt{a^3/(GM)}$. The mean motion is then $n = 2\pi/P$, and the epoch of periastron passage follows from this as $t_\mathrm{peri} = t - \mathrm{MA}/n - t_\mathrm{ref}$. Finally,
\begin{equation}
    \tau = \frac{t_{\mathrm{peri}}}{P}.
\end{equation}
The corner plot in Figure \ref{fig:pairplot-octo-tau-theta} demonstrates the benefit of this parameterization.
The bottom rows show the same orbits parameterized with $\tau$ and with $\theta$.
The $\theta$ marginal posterior is nearly Gaussian, and has simple relationships with all other orbital parameters.
By contrast, the $\tau$ marginal posterior is less informative and has complex relationships to $\omega$ and $\Omega$.

\section{Thiele-Innes Elements \label{sec:alternate-parameterization}}

In Appendix \ref{sec:cartesian-coordinates}, we saw that Keplerian orbits can be uniquely characterized by the Campbell elements $(a, e, i, \omega, \Omega)$, in addition to an orbit position parameter ($t_{\mathrm{peri}}$, $\tau$, or $\theta$).
However, Keplerian orbits can be equivalently characterized by the Thiele-Innes elements $(e, A, B, F, G)$, where $e$ is the orbital eccentricity and $A$, $B$, $F$, and $G$ are the Thiele-Innes constants, along with an orbit position parameter.
The Thiele-Innes constants are defined as
\begin{align}
    A &= a\bar{\omega}(\cos(\Omega)\cos(\omega) - \sin(\Omega)\sin(\omega)\cos(i)), \label{eq:thiele-innes-A}\\
    B &= a\bar{\omega}(\sin(\Omega)\cos(\omega) + \cos(\Omega)\sin(\omega)\cos(i)), \label{eq:thiele-innes-B}\\
    F &= a\bar{\omega}(-\cos(\Omega)\sin(\omega) - \sin(\Omega)\cos(\omega)\cos(i)), \label{eq:thiele-innes-F}\\
    G &= a\bar{\omega}(-\sin(\Omega)\sin(\omega) + \cos(\Omega)\cos(\omega)\cos(i)), \label{eq:thiele-innes-G}
\end{align}
where $\bar{\omega}$ is the parallax distance to the system. 
As we can see, it is straightforward to calculate $A$, $B$, $F$, and $G$ from $a$, $i$, $\omega$, and $\Omega$. 
Inverting this relationship, although less straightforward, is still doable. We begin by defining
\begin{align}
    u &= \frac{1}{2}(A^2 + B^2 + C^2 + D^2), \\
    v &= AG - BF.
\end{align}
We can recover $a$ using $u$ and $v$ like so:
\begin{equation}
    a = \frac{\sqrt{u + \sqrt{(u + v)(u - v)}}}{\bar{\omega}}. \label{eq:a-from-uv}
\end{equation}
Once we have $a$, $i$ immediately follows from
\begin{equation}
    i = \arccos\left( \frac{v}{a^2\bar{\omega}^2} \right).
\end{equation}
Since Eqs. \ref{eq:thiele-innes-A} -- \ref{eq:thiele-innes-G} involve only $\cos(i)$, and $\cos(\theta) = \cos(-\theta)$, this is technically an equation for $|i|$. 
Next, we define 
\begin{align}
    j &= \arctan\left(A + G, B - F \right), \\
    k &= \arctan\left( A - G, B + F \right),
\end{align}
where $\arctan(y,x)$ is a quadrant-sensitive version of $\arctan(y/x)$ (sometimes written as $\mathrm{arctan2}(y,x)$ or $\mathrm{atan2}(y,x)$). 
From these quantities, $\omega$ and $\Omega$ are given by
\begin{align}
    \omega &= \frac{1}{2}(j - k), \\
    \Omega &= \frac{1}{2}(j + k).
\end{align}
Thus, we have successfully recovered $a$, $i$, $\omega$, and $\Omega$ from the Thiele-Innes constants. 





\bibliography{bib-extras,zotero}
\bibliographystyle{aasjournal}



\end{document}